\documentclass[structabstract]{aa}  

\usepackage{amssymb}
\usepackage{amsmath}
\usepackage{graphicx,subfigure}
\usepackage{hyperref}
\hypersetup{breaklinks=true}
\usepackage{breakurl} 
\usepackage{pdflscape}
\usepackage{afterpage}
\usepackage{capt-of}
\usepackage{rotating}

\usepackage[utf8]{inputenc}
\usepackage{natbib}

\begin{document}

   \title{Improved angular momentum evolution model for solar-like stars}

   \subtitle{II. Exploring the mass dependence}

   \author{F. Gallet\inst{1,2}          
          \and
          J. Bouvier\inst{1,2}
          }

   \institute{
 Univ. Grenoble Alpes, IPAG, F-38000 Grenoble, France\and 
CNRS, IPAG, F-38000 Grenoble, France   \\
   \email{florian.gallet@obs.ujf-grenoble.fr}}

   \date{Received --; accepted --} 

\abstract{Understanding the physical processes that dictate the angular momentum evolution of solar-type stars from birth to maturity remains a challenge for stellar physics.}{We aim to account for the observed rotational evolution of low-mass stars over the age range from 1 Myr to 10 Gyr. }{We developed angular momentum evolution models for 0.5 and 0.8 $M_{\odot}$ stars.  The parametric models include a new wind braking law based on recent numerical simulations of magnetised stellar winds, specific dynamo and mass-loss rate prescriptions, as well as core-envelope decoupling. We compare model predictions to the distributions of rotational periods measured for low-mass stars belonging to star-forming regions and young open clusters. Furthermore, we explore the mass dependence of model parameters by comparing these new models to the solar-mass models we developed earlier.}{Rotational evolution models are computed for slow, median, and fast rotators at each stellar mass. The models reproduce reasonably well the rotational behaviour of low-mass stars between 1~Myr and 8-10~Gyr, including pre-main sequence to zero-age main sequence spin up, prompt zero-age main sequence spin down, and early-main sequence convergence of the surface rotation rates. Fast rotators are found to have systematically shorter disk lifetimes than moderate and slow rotators, thus enabling dramatic pre-main sequence spin up. They also have shorter core-envelope coupling timescales, i.e., more uniform internal rotation. As for the mass dependence, lower mass stars require significantly longer core-envelope coupling timescales than solar-type stars, which results in strong differential rotation developing in the stellar interior on the early main sequence. Lower mass stars also require a weaker {braking torque} to account for their longer spin-down timescale on the early main sequence, while they ultimately converge towards lower rotational velocities than solar-type stars in the longer term because of their reduced moment of inertia. We also find evidence that the mass dependence of the wind braking efficiency may be related to a change in the magnetic topology in lower mass stars.}
 {We have included in parametric models the main physical processes that dictate the angular momentum evolution of low-mass stars. The models suggest that these processes are quite sensitive to both mass and instantaneous rotation rate. We have worked out and reported here the main trends of these mass and rotation dependencies, whose origin still have to be addressed through a detailed modelling of magnetised stellar winds, internal angular momentum transport processes, and protoplanetary disk dissipation mechanisms.}

   \maketitle
%
\defcitealias{Gallet13}{Paper I} 	

\section{Introduction}


Observational constraints on the rotational evolution of low-mass stars have exponentially increased in the last decade, thanks to a number of dedicated long-term monitoring studies of nearby populations \citep[see e.g.][]{Irwin09a,Hartman10,Agueros11,Meibom2011,Irwin11,Affer2012,Affer2013,PPVI,Gallet13, McQuillan2014}. These new observational results provide extremely useful guidance for the modelling of angular momentum evolution of low-mass stars  ($M_* < 1.2 M_{\odot}$) from 1~Myr to 10~Gyr   \citep[e.g.][]{Irwin07,Bouvier08,Den10,Spada11,Reiners2012,Gallet13} and offer a unique insight into the physical processes that dictate rotational evolution. To account for the observations, parametric models have to incorporate at least three major processes: the star-disk interaction during the early pre-main sequence (PMS) \citep{MP05a,Zanni09, FPA00,Matt10,Zanni2012}, the loss of angular momentum through powerful stellar winds on the early main sequence (MS);\citep{WD67, Kawaler88, MP05b,Vidotto2011, Zanni11,Cranmer11,Matt11a, Reiners2012,Matt12, Reville14, Vidotto14b}, and the redistribution of angular momentum in the stellar interior at all phases of evolution \citep{McGB91, Allain98, Talon03,Palacios03, Talon05, Palacios06,Charbonnel05, Charbonnel10, Lagarde11,Lagarde12, Charbonnel13}. The angular velocity evolution of low-mass stars appears to be controlled to a large extent by these three physical processes, from their birth to the end of their MS evolution and possibly beyond.

In the first paper of this series \citep{Gallet13}, we developed a model to account for the angular momentum evolution of solar-mass stars. The parametric model included star-disk locking, a new braking law with dynamo, and mass-loss prescriptions that relied on the latest numerical simulations \citep{Cranmer11, Matt12}, and core-envelope decoupling to account for the reduced efficiency of internal angular momentum transport processes.  The aim of the present study is to extend this model to lower mass stars: Can similar parametric models account for the evolution of lower mass stars? How do model parameters depend on mass?  What does this mass dependence reveal about the underlying physical processes? These are the questions we attempt to address in this study.
In Sect.~\ref{dataset}, we present the set of 18 rotational period distributions that we used to constrain our models at various ages.  We briefly describe in Sect. \ref{modcons} the assumptions we made to compute the angular momentum evolution of low-mass stars from birth to the end of the MS. The results are presented in Sect.~\ref{res}. In Sect.~\ref{disc}, we work out the mass dependence of angular momentum evolution and discuss its implication for the physical mechanisms involved. We will also discuss the impact of these new models on gyrochronology and lithium depletion. Conclusions are drawn in Sect.~\ref{conc}.

\section{Observational constraints on the rotational evolution of low-mass stars: the data sets}
\label{dataset}

The goal of the angular momentum evolution models is to reproduce the observed rotational evolution of low-mass stars from birth to the end of the main sequence. To characterise the latter, we used rotational period distributions measured for the coeval populations of 18 star-forming regions and open clusters, thus spanning an age range from 1 Myr to 1 Gyr, to which we added results pertaining to old disk field stars from \citet{McQuillan2014}. The data for star-forming regions and young clusters, including their age and angular velocity percentiles (25$^{th}$, 50$^{th}$, and 90$^{th}$) in each mass bin investigated here (0.4-0.6, 0.7-0.9, and 0.9-1.1 M$_\odot$), are provided in Table \ref{opencluster} \citep[see Appendix A of][for a detailed description of the properties of these clusters]{Gallet13}.

\begin{sidewaystable*}
\vspace{+18.45cm} 
\caption{Star-forming regions and open clusters whose rotational distributions are used is this study.}   
\label{opencluster}      
\centering   
\small
\begin{tabular}{l l l c c c c c c c c c c c}      
\hline\hline                
Cluster & Age & Ref.  & -- & $\Omega_{25}$ & -- & -- & $\Omega_{50}$ &-- & -- & $\Omega_{90}$ & --   \\  
 &(Myr)& & $1~M_{\odot}$ & $0.8~M_{\odot}$ & $0.5~M_{\odot}$  & $1~M_{\odot}$ & $0.8~M_{\odot}$ & $0.5~M_{\odot}$  & $1~M_{\odot}$ & $0.8~M_{\odot}$ & $0.5~M_{\odot}$  \\
\hline   
ONC	 & 	1.5	 & 	1	 & 	3.28$\pm$0.3	 & 	2.86$\pm$0.14	 & 	3.59$\pm$0.26	 & 	4.58$\pm$0.86	 & 	4.03$\pm$0.58	 & 	5.27$\pm$0.89	 & 	13.21$\pm$2.14	 & 	20.11$\pm$1.93	 & 	14.93$\pm$3.63	  	\\
NGC 6530	 & 	2	 & 	2	 & 	3.87$\pm$0.55	 & 	2.89$\pm$0.19	 & 	3.19$\pm$0.21	 & 	6.15$\pm$0.9	 & 	4.08$\pm$1.14	 & 	5.74$\pm$0.54	 & 	22.32$\pm$5.15	 & 	18.73$\pm$2.73	 & 	22.72$\pm$3.48		\\
NGC 2264	 & 	3	 & 	3	 & 	--	 & 	2.81$\pm$0.5	 & 2.81$\pm$0.38	& --	 & 	5.32$\pm$0.91	 & 	6.16$\pm$1.93	 &	--	 & 	10.66$\pm$2.01	 & 	27.57$\pm$2.83	 	\\
Cep OB3b	 & 	4	 & 	4	 & 	3.25$\pm$0.39	 & 	3.91$\pm$0.81	 & 	3.38$\pm$0.2	 & 	6.99$\pm$1.07	 & 	6.45$\pm$1	 & 	5.58$\pm$0.43	 & 	14.98$\pm$1.55	 & 	17.13$\pm$2.14	 & 	23.66$\pm$5.42		\\
NGC 2362	 & 	5	 & 	5	 & 	3.13$\pm$0.29	& 2.75$\pm$0.29	& 3.9$\pm$0.26	 & 4.2$\pm$0.58	 & 3.82$\pm$0.29	& 6.31$\pm$0.67	 & 	12.36$\pm$4.76	 & 	14.01$\pm$4.4	 & 	23.09$\pm$5.94	 	\\
h Per & 13& 6	 & 	4.75$\pm$0.19	 & 3.93$\pm$0.17 & 4.5$\pm$0.53	 & 8.31$\pm$0.85 &	6.33$\pm$0.74 & 11.93$\pm$5.03	 &  73.37$\pm$6.32	 & 	46.99$\pm$6.2	 & 	58.4$\pm$4.37	  	\\
NGC 2547	 & 	35 & 	7	 & --& 4.55$\pm$0.38	 & 	4.77$\pm$0.87	 & 	--	 & 5.28$\pm$1.01 & 10.82$\pm$1.86	 & --	 & 	47.31$\pm$22.35	 & 	45.24$\pm$17.7 \\	
IC 2391	 & 	50	 & 	8	 & --  & 4.38$\pm$0.76	 & 	--	 & 			-- 	 & 	9.76$\pm$4.29	 & 	--	 & 	 	--	  			 & 	98.13$\pm$31.77	 & 	-- \\	
$\alpha$ Per	 & 	80	 & 	8	 & 	-- 	 & 5.88$\pm$0.75	 &	--	 & --	 & 	14.81$\pm$8.32	 & 	--	 & 	-- & 	98.41$\pm$14.26	 & 	-- \\		
Pleiades	 & 	125	 & 	9	 & 	4.92$\pm$0.13	 & 	3.59$\pm$0.08	 & 	7.98$\pm$1.2	 & 	6.29$\pm$0.31	 & 	4.27$\pm$0.24	 & 	18.24$\pm$1.91	 & 	37.99$\pm$10.08	 & 	61.49$\pm$5.5	 & 	76.74$\pm$4.91 \\		
M 50	 & 	130	 & 	10	 & 	3.38$\pm$0.3	 & 	3.21$\pm$0.1	 & 	6.79$\pm$0.53	 & 	5.17$\pm$0.36	 & 	4.23$\pm$0.22	 & 	17.19$\pm$1.74	 & 	15.41$\pm$5.16	 & 	43.12$\pm$3.55	 & 	72.22$\pm$3.32 \\	 	
M 35	 & 	150	 & 	11	 & 	4.39$\pm$0.07	 & 	3.38$\pm$0.1	 & --	 & 5.05$\pm$0.22	 & 	4.59$\pm$0.71	 & 	--	 & 24.3$\pm$5.27	 & 59.3$\pm$5.55	 & 	-- \\	  	
NGC 2516	 & 	150	 & 	12	 & 	--	 & 			--	 & 			5.11$\pm$0.66	 & 	--	 & 	--	 & 	 	11.52$\pm$1.56	 & 	--	 & 	--		 & 	70.12$\pm$4.39 \\		
M 34	 & 	220	 & 	13	 & 	3.68$\pm$0.14	 & 	2.67$\pm$0.1	 & 	--	 & 4.64$\pm$0.83	 & 	3.07$\pm$0.09	 & 	--	 & 27.02$\pm$6.78	 & 	21.31$\pm$5.14	 & 	-- \\	
M 37	 & 550 & 14 & 2.95$\pm$0.05 & 2.38$\pm$0.04	 & 1.97$\pm$0.44 & 3.26$\pm$0.06 & 2.63$\pm$0.06 & 11.23$\pm$3.37	 & 3.75$\pm$0.13 & 	3.7$\pm$0.27	 & 40.69$\pm$2.14 \\		
Praesepe	 & 580	 & 	15	 & 	2.73$\pm$0.05	 & 	2.11$\pm$0.05	 & 	1.39$\pm$0.07	 & 	2.85$\pm$0.02	 & 	2.52$\pm$0.11	 & 	3.28$\pm$1.57	 & 	2.92$\pm$0.75	 & 	5.94$\pm$1.39	 & 	22.03$\pm$4.63 \\	 	
Hyades	 & 	625	 & 	16	 & 	2.44$\pm$0.03	 & 	2.05$\pm$0.04	 & 	1.41$\pm$0.07	 & 	2.62$\pm$0.06	 & 	2.18$\pm$0.03	 & 	1.91$\pm$0.09	 & 	2.91$\pm$0.1	 & 	2.46$\pm$0.15	 & 	2.3$\pm$0.78 \\	 	
NGC 6811	 & 	1000	 & 	17	 & 	2.27$\pm$0.04	 & 	--	 & --	 & 2.36$\pm$0.01	 & 	--	 & 	 	--	 & 	2.45$\pm$0.05	 & 	--	 & 	-- \\	 
\hline   
\end{tabular}
\tablebib{(1)~\citet{RL2009}; (2) \citet{Henderson11}; 
(3) \citet{Cieza07}; (4) \citet{Littlefair10}; (5) \citet{Irwin08a}; (6) \citet{Moraux13}; (7) \citet{Irwin08b}; (8) Irwin et Bouvier; (9) \citet{Hartman10}; (10) \citet{Irwin09b}; (11) \citet{Meibom09};  (12) \citet{Irwin07};  (13) \citet{Meibom2011b}; (14) \citet{Hartman09}; (15) \citet{Agueros11}+\citet{Delorme11}; (16) \citet{Delorme11}; (17) \citet{Meibom2011}.}
\end{sidewaystable*}

\begin{figure*}[ht!]
     \begin{center}
    \includegraphics[angle=-90,width=15cm]{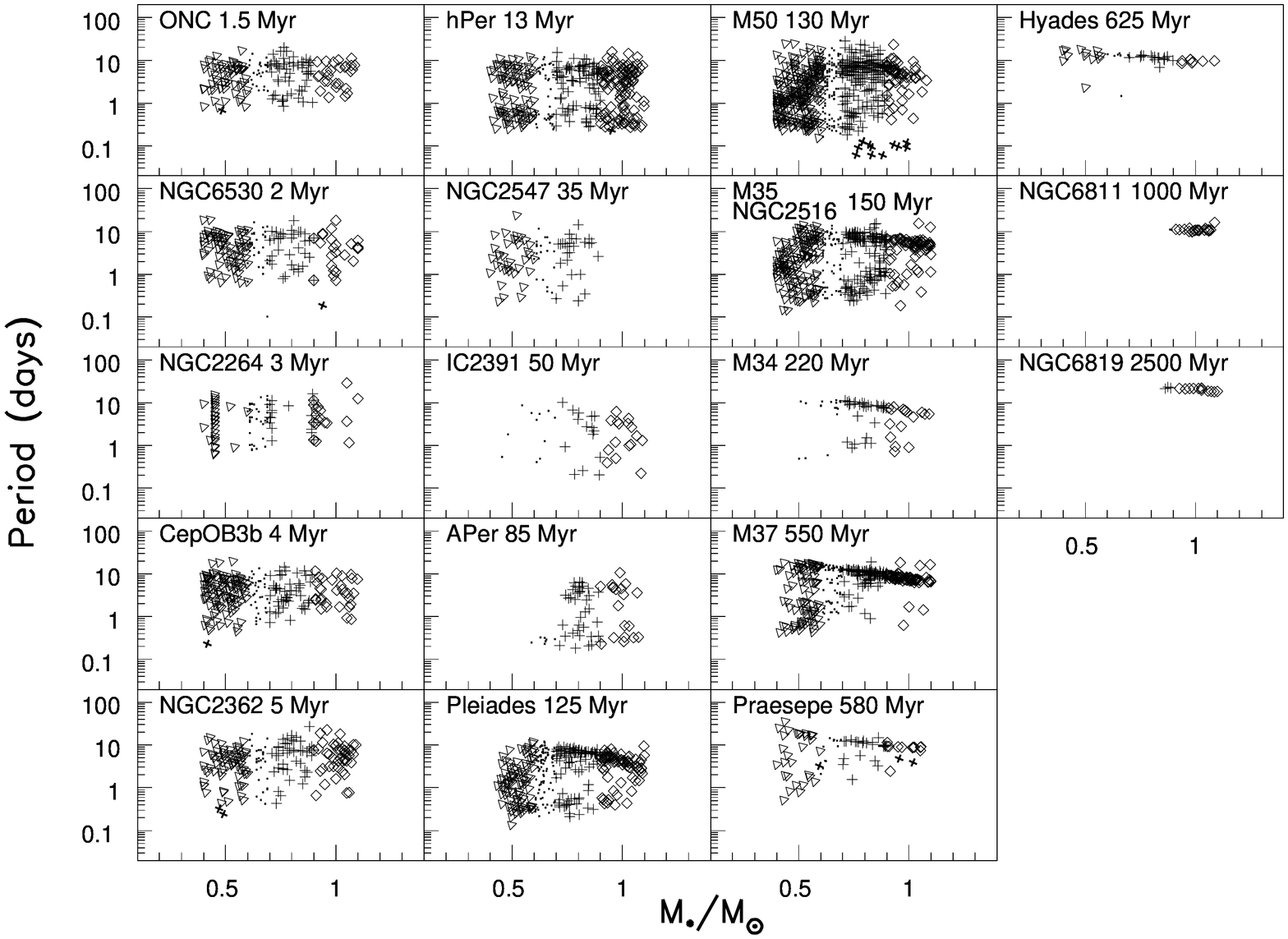}
    \end{center}
    \caption{The rotational period distribution of 0.4-0.6, 0.7-0.9, and 0.9-1.1 M$_\odot$ stars (respectively shown as tilted triangles, crosses, and tilted squares) in star-forming regions and young open clusters is plotted as a function of stellar mass. The panels are ordered by increasing age, from top to bottom and left to right. The few black tilted crosses in NGC 6530, M50, and Praesepe are stars rejected from the samples for reasons explained in the text. The rotational periods of stars outside the selected mass bins are shown as small dots, as are subsamples not rich enough (i.e. $N_{star} < 15$) to yield statistically meaningful percentiles and therefore rejected from the analysis. }
   \label{protmass}
\end{figure*}

Figure \ref{protmass} shows the distributions of rotation periods plotted as a function of stellar mass for star-forming regions and young open clusters. Several patterns can be recognised in this figure that characterise the rotational evolution of young low-mass stars. The youngest star-forming regions, with an age less than 10-15 Myr, exhibit a wide distribution of rotation periods, ranging from less than 1 and about 10 days, with little dependence on mass over the mass range investigated here (0.4-1.1 M$_\odot$). A well-defined relationship between rotation and mass starts to be seen in older clusters, from the ZAMS onwards, which first appears for solar-mass stars and later propagates to lower masses. The oldest clusters indeed exhibit quite a tight rotation-mass correlation, with rotation rates steadily decreasing towards lower masses. It is also noteworthy that the dispersion of rotational velocities is the lowest at ZAMS for very low-mass stars (cf. M50, 130 Myr) and increases later-on on the early MS over a few 100 Myr, while it is quite the opposite for solar-type stars whose wide rotational dispersion at ZAMS (cf. Pleiades, 125 Myr) is promptly erased on the early MS. This complex behaviour reflects the combination of various physical processes acting on the angular momentum content of the stars as they evolve.

\begin{figure}[ht!]      \begin{center}             \includegraphics[angle=-90,width=0.4\textwidth]{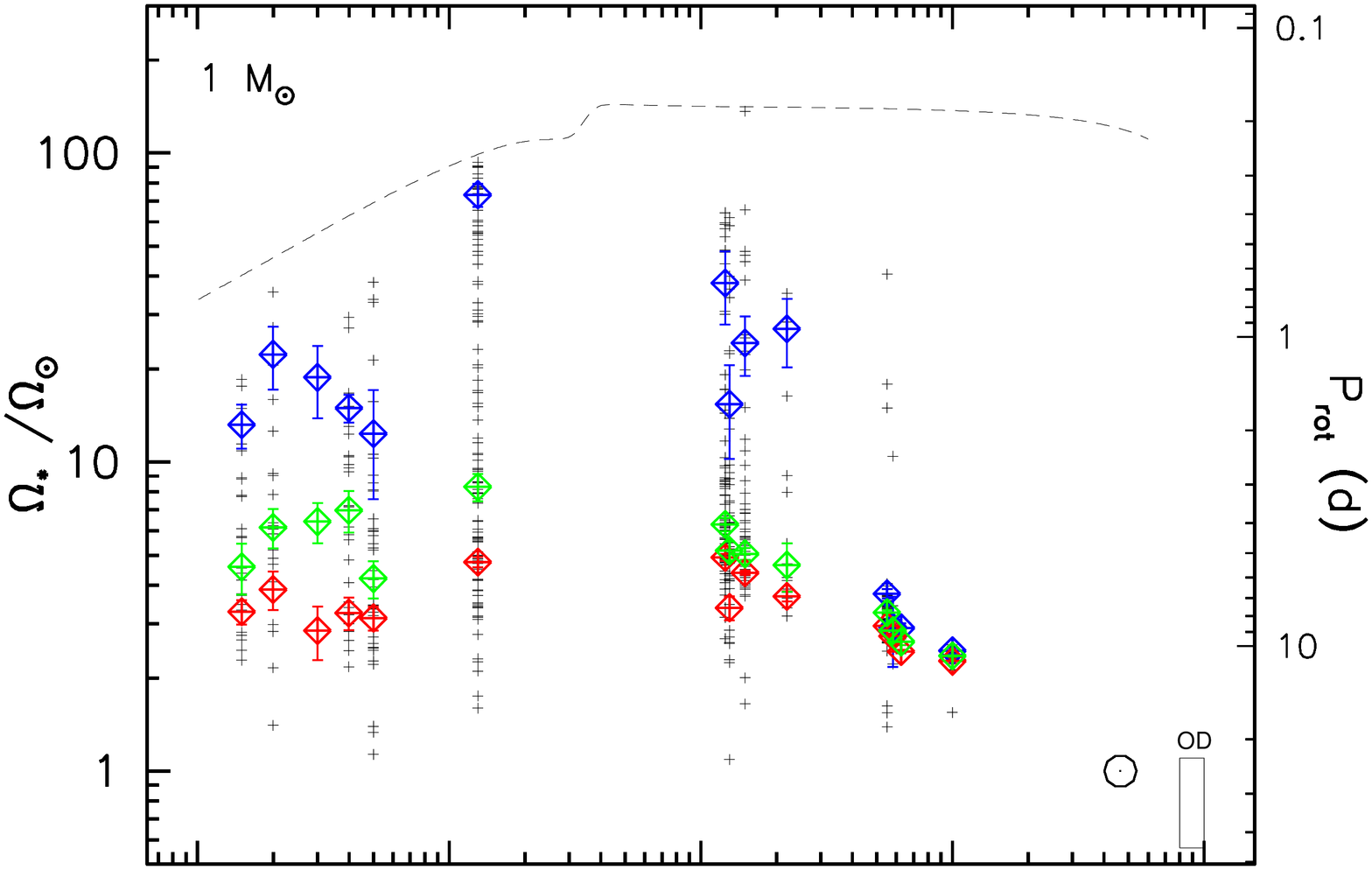}\\             \vspace{-0.9cm}            \includegraphics[angle=-90,width=0.4\textwidth]{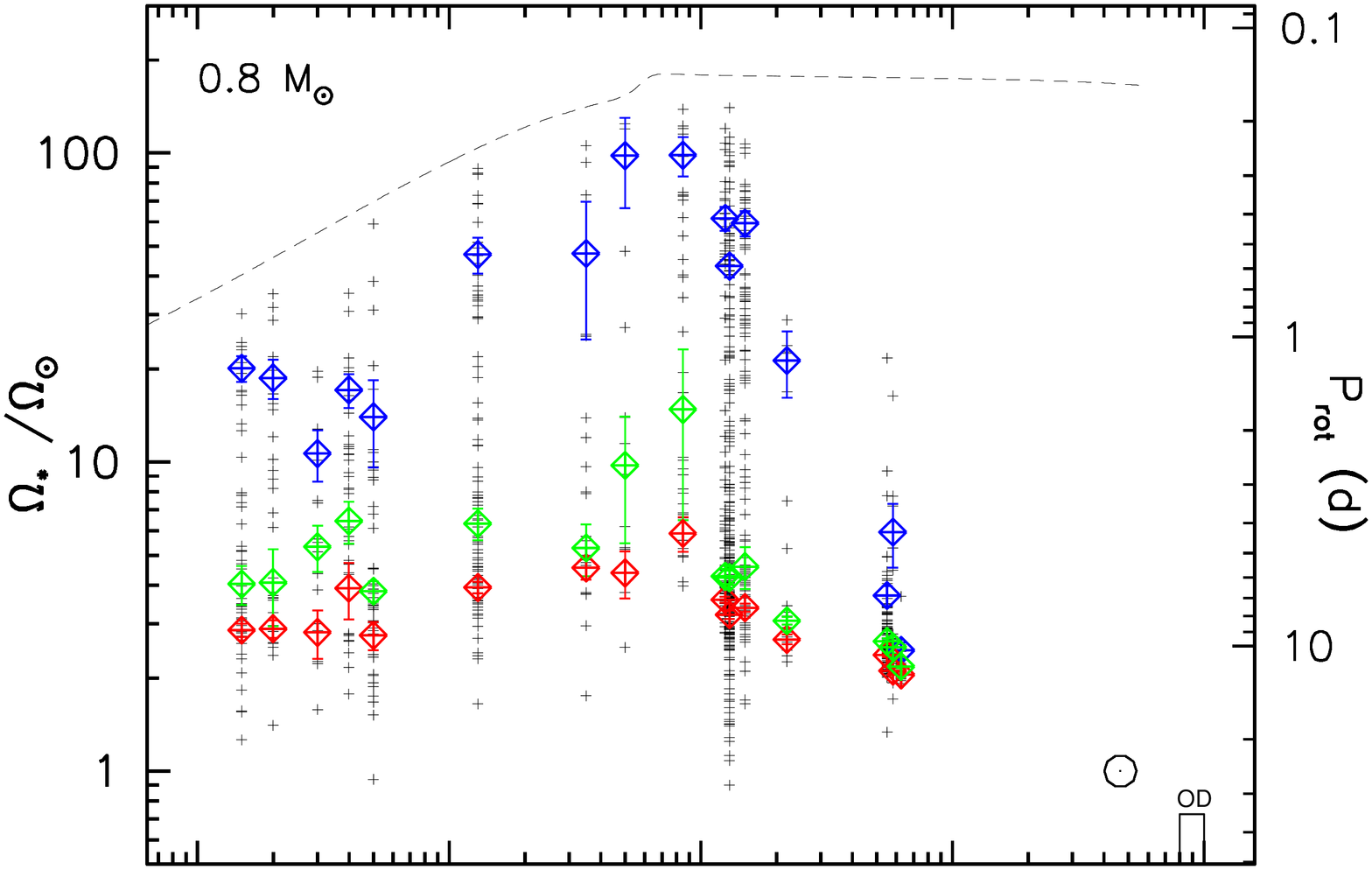}\\            \vspace{-0.9cm}             \includegraphics[angle=-90,width=0.4\textwidth]{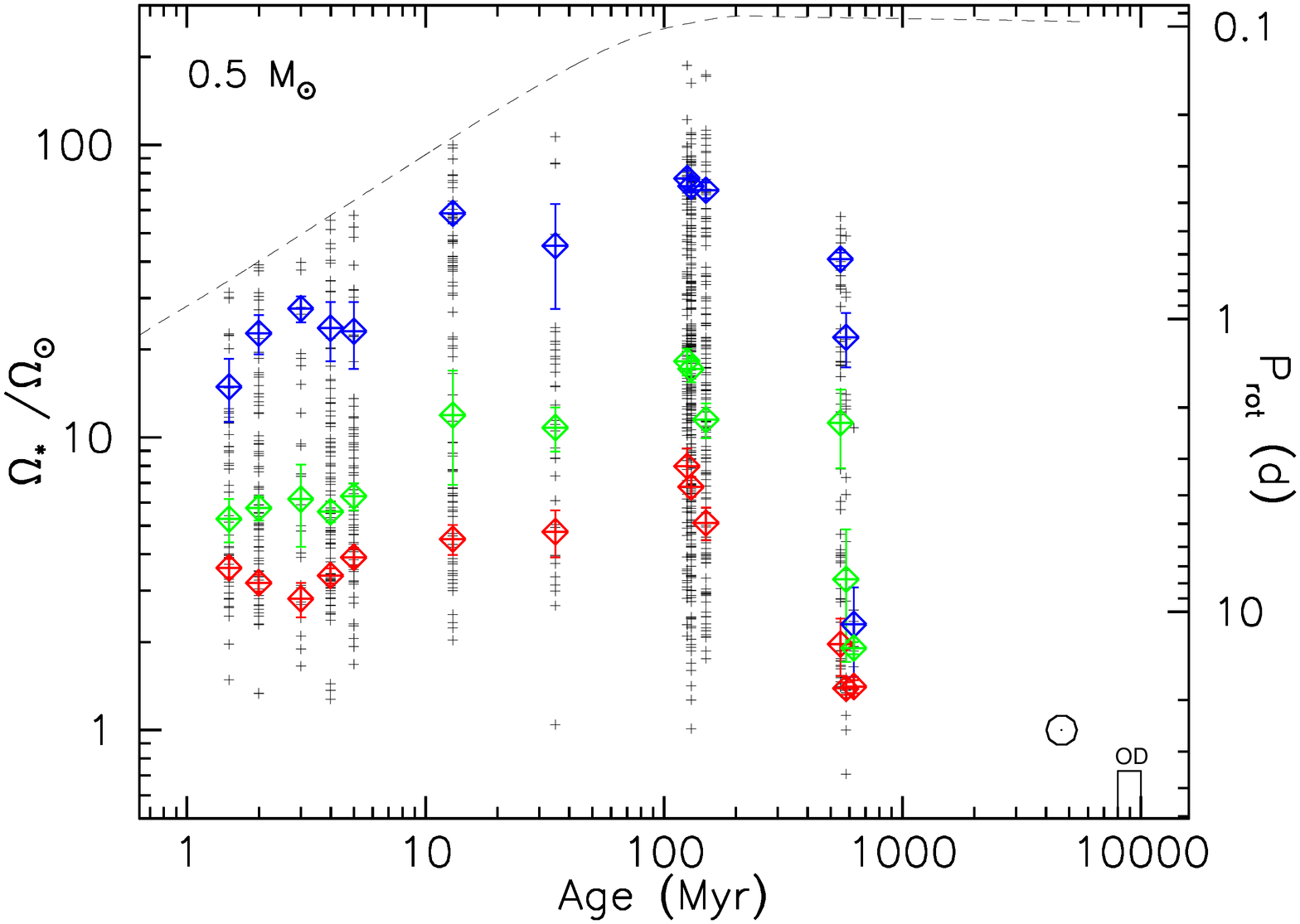}     \end{center}     \caption{     Angular velocity distributions are plotted as a function of time for low-mass stars in star-forming regions and young open clusters. Each panel covers a different mass bin:  0.9-1.1 $ M_{\odot}$ (\textit{upper panel}), 0.7-0.9 $M_{\odot}$ (\textit{middle panel}), 0.4-0.6 $M_{\odot}$ (\textit{lower panel}). The red, green, and blue tilted squares and associated error bars represent the 25$^{th}$, 50$^{th}$, and 90$^{th}$ rotational percentiles, respectively. The open circle shows the angular velocity of the present Sun for reference. The black rectangle labelled OD (lower right corner of each panel) shows the angular velocity dispersion of old disk field stars. The black dashed line represents the evolution of break-up velocity. The left vertical axis is labelled with angular velocities normalised to the Sun's, while the right vertical axis is labelled with rotational periods (days).}    
\label{compamas} \end{figure}

Figure \ref{compamas} shows the same data set as in Fig.~\ref{protmass} except that for each cluster, we extracted the observed angular velocity distribution in a given mass bin and plotted it as a function of time. The three panels of Fig. \ref{protmass} thus illustrate the evolution of the rotational distributions of 0.5, 0.8, and 1 M$_\odot$ stars, respectively. The $25^{th}$, $50^{th}$, and $90^{th}$ percentiles of the distributions and their respective error bars were computed at each mass and age as described in \citet{Gallet13}. Figure \ref{compamas} suggests a qualitatively similar evolution of the rotation rates for stars in the three mass bins. The evolution is characterised by a spin rate that hardly or slowly varies during the first few Myr, then followed by a rapid acceleration towards the ZAMS, and a subsequent braking over longer timescales on the MS. Quantitatively, the rotational evolution of the lowest mass stars appears delayed compared to solar-type stars, as they reach the ZAMS later and are braked on the main sequence over longer timescales. The models presented in the next sections aim to reproduce the overall evolution as well as subtle differences as a function of mass.

While these distributions appear statistically robust and relatively unbiased, some residual contamination by non-cluster members cannot be totally dismissed. Indeed, some stars appear to rotate beyond break-up, presumably due to contamination of the samples by short period field binaries\footnote{We thus rejected M50-5-1624, -3-1468, -3-464, -7-7623, -3-5840, -7-5624, -5-2673, -3-7334, -3-2531, -6-1574, -4-2077, -4-4939, and -8-6076 from M50 \citep{Irwin09b}, N2362-2-6989, and N2362-5-4947 from NGC~2362, \#11041 \citep[number from][]{Hillenbrand97} from ONC, and \#494 \citep[number from][]{Moraux13} from h Per, and \#3245 \citep[number from][]{Littlefair10} from CepOB3b}. The reported photometric period is sometimes associated with harmonic of the rotation period e.g. in case of two spots located at opposite longitudes on the stellar surface\footnote{We thus rejected 1SWASP J083722.23+201037.0 (KW 30) and 1SWASP J084005.72+190130.7 (KW 256) in Praesepe \citep{Delorme11}.}. Finally, some stars simply have a relatively low membership probability\footnote{We thus rejected JS 634 in Praesepe that has a membership probability of 0.62, compared to 0.94 for cluster members \citep{Agueros11}.} or an age estimate in conflict with that of the cluster\footnote{In NGC6530 (2 Myr),  we rejected XID 138 (0.94 $ M_{\odot}$, $P_{rot} = 0.18743$ d) as its age is estimated to be 22 Myr and is probably already accelerating towards the ZAMS \citep{Henderson11}.}.

Another issue lies in the difficulty of estimating the age of clusters. Thus, for very young clusters with an age less than about 5-10 Myr, age uncertainties may be approximately the actual age estimate \citep{Bell13}. Fortunately, the change in rotation rates at these very young ages are relatively mild, so that the modelling does not suffer much from these uncertainties (i.e. shuffling the youngest clusters within the 1-5 Myr age range in Fig.~\ref{compamas} would not significantly impact the models). The ages of older clusters is much better known, with uncertainties close to 10\%. We therefore neglect the age uncertainties in the models developed below.

\section{Parametric models of angular momentum evolution: the assumptions}
\label{modcons}
The angular momentum evolution of isolated low-mass stars is controlled by the balance between three main physical mechanisms: angular momentum removal by magnetised stellar winds (here after ``the wind braking''), the star-disk interaction, and the angular momentum transfer within the stellar interior. We discuss in this section the corresponding model assumption. 

\subsection{Structural stellar evolution}

We adopt the \citet{Baraffe98} non-rotating, solar-metallicity models computed for 0.5, 0.8, and 1 M$_{\odot}$ stars, with a mixing length parameter $\alpha$=1.9 and an helium abundance $Y=0.275$. These models provide the structural evolution (i.e. mass, radius, moment of inertia) of the inner radiative core and outer convective envelope from the early PMS to the end of the MS i.e. from 1 Myr to 10 Gyr.

\begin{figure}
\centering
\includegraphics[angle=-90,width=9cm]{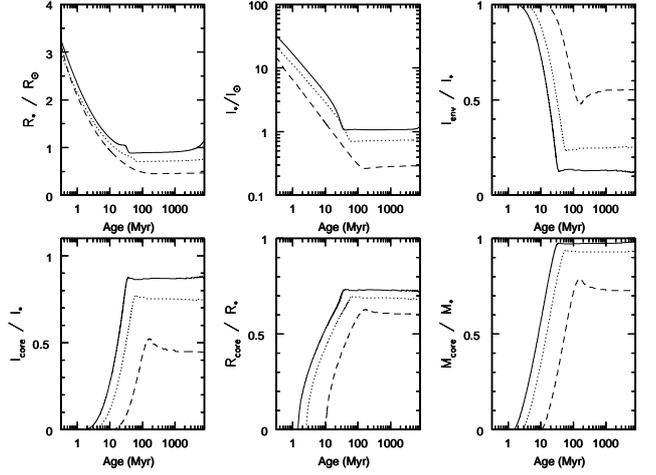}
\caption{Evolution of the main stellar parameters from the \citet{Baraffe98}'s models for 1 $M_{\odot}$ (solid line), 0.8 $M_{\odot}$ (dotted line), and 0.5 $M_{\odot}$ (dashed line). Upper panel, from left to right: stellar radius (in solar unit), moment of inertia (in solar unit, $I_{\odot} = 6.411 \times 10^{53} g.cm^{-2}$), and moment of inertia of the convective envelope (in stellar unit). Lower panel, from left to right: the radiative core's moment of inertia, radius, and mass (in stellar unit). }
\label{stellarparam}
\end{figure}

Figure \ref{stellarparam} shows the evolution of the internal structure  for 0.5, 0.8, and 1 M$_\odot$ stars. Most stellar quantities increase for higher mass stars (i.e. the stellar radius and moment of inertia, the moment of inertia of the radiative core, as well as its radius and mass), except for the moment of inertia of the convective envelope that decreases for higher masses.  The size and mass of the radiative core is strongly mass dependent. While for a 1 $M_{\odot}$ star the radiative core represents about 73\% of the stellar size, this fraction falls to 60\% for a 0.5 $M_{\odot}$ star. Similarly, for a 1 $M_{\odot}$ star almost all the stellar mass is contained in the radiative core while only 73\% of the total mass is stored in the core of a 0.5 $M_{\odot}$ star. Note that the rapid decrease of the star's moment of inertia during PMS evolution is expected to strongly impact on the star's surface velocity, since angular momentum is given by $J = I\cdot\Omega$ where $I$ is the moment of inertia and $\Omega$ the angular velocity.

\subsection{Magnetised stellar winds }
\label{SW}

Assuming a spherical outflow, the braking torque exerted by a magnetised stellar wind on the stellar surface can be expressed as
\begin{eqnarray}
\label{djdt}
\Gamma_{wind} \propto \Omega_* \cdot \dot{M}_{wind} \cdot r_A^2,
\end{eqnarray}
where $\dot{M}_{wind}$ is the mass-loss rate, and $r_A$ is the averaged value of the Alfv\'en radius. The latter is obtained from \citet{Matt12} as

\begin{eqnarray}
\label{ranew}
r_A = K_1 \left[ \frac{B_p^2 R_*^2}{\dot{M}_{wind} \sqrt{K_2^2v_{esc}^2 + \Omega_*^2 R_*^2}}\right]^m R_*,
\end{eqnarray}
where $B_p$ is the surface strength of the dipole magnetic field at the stellar equator, $v_{esc}=\sqrt{2GM_*/R_*}$ is the escape velocity, and m = 0.22, $K_2$= 0.0506. As in \citet{Gallet13}, we identify $B_p$ to the strength of the mean magnetic field $B_*f_*$where $B_*$ is the magnetic field intensity and $f_*$ is the filling factor, i.e. the fraction of the stellar surface that is magnetised. The evolution of the mean magnetic field $B_*f_*$ as a function of stellar density, effective temperature, and angular velocity is taken from the model developed by \citet{Cranmer11}. To reproduce the mean solar filling factor ($f_{\odot}=10^{-1}-10^{-3}$), we slightly modified the expression of $f_*$ given by \citet{Cranmer11}, as described in \citet{Gallet13}. Eventually, this yields
\begin{eqnarray}
 B_*f_* &=& 1.13 \times B_{eq} \times f_*, \nonumber \\
 &=& 1.13 \sqrt{\frac{8 \pi \rho_* k_B T_{eff}}{\mu m_H}} \frac{0.55}{\left[1 + (x/0.16)^{2.3}\right]^{1.22}},
 \label{bf}
\end{eqnarray}
where $x = Ro / Ro_{\odot}$ and $Ro = P_{rot}/\tau_{conv}$ is the Rossby number, i.e. the ratio of   $P_{rot}$ the rotation period to $\tau_{conv}$ the convective turnover timescale, with $Ro_{\odot}=1.96$ \citep{Wright11,Oglethorpe13}. In Eq.~\ref{bf}, {$B_{eq}$ is the equipartition magnetic field strength \citep{Cranmer11}} with $\rho_*$ the photospheric mass density, $k_B$ the Boltzmann's constant, $T_{eff}$ the effective temperature, $\mu$ the mean atomic weight, and $m_H$ the mass of a hydrogen atom.

The mass-loss rate prescription we inject in Eq.~\ref{ranew} also stems from \citet{Cranmer11} coronal wind models
\begin{eqnarray}
\dot{M}_{wind} \propto \left(\frac{R_*}{R_{\odot}}\right)^{16/7} \left(\frac{L_*}{L_{\odot}}\right)^{-2/7} \left(F_{A*}\right)^{12/7}  f_*^{5/7},
\label{mdot}
\end{eqnarray}
where $L_*$ is the stellar luminosity, and $F_{A*}$ is the quantity of energy deposited in the photosphere by Alfv\'en waves. This energy is converted into a heat energy flux that powers the stellar wind. We also modified the original expression of $F_{A*}$ from \citet{Cranmer11} to incorporate the difference of mixing length parameters used in their study and ours  (see details in  \citet{Gallet13}).

\begin{figure*}
\centering
\includegraphics[angle=-90,width=12cm]{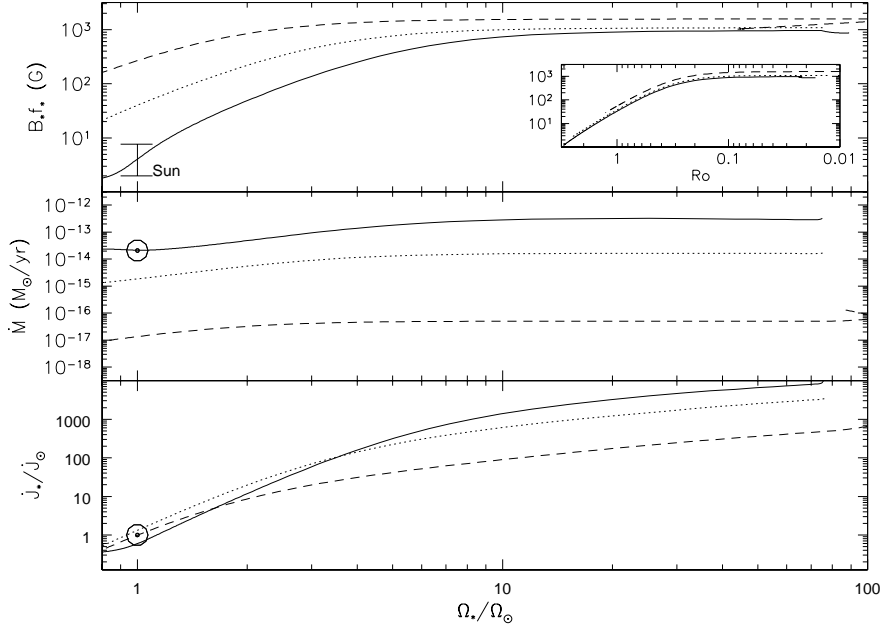}
\caption{\textit{Upper panel:} Mean magnetic field strength as a function of stellar angular velocity  for 1, 0.8, and 0.5 $M_{\odot}$ stars (solid, dotted, and dashed lines, respectively). The insert shows the mean magnetic field as a function of the Rossby number. The Sun's range of $B_*f_* = 2-7.7~G$ is shown as a vertical bar. \textit{Middle panel:} Mass-loss rate as a function of stellar angular velocity  for 1, 0.8, and 0.5 $M_{\odot}$ stars (solid, dotted, and dashed lines, respectively). The Sun's present day mass-loss is shown as an open circle. \textit{Lower panel:} Angular momentum loss rate normalised to the Sun's as a function of angular velocity for 1, 0.8, and 0.5 $M_{\odot}$ stars (solid, dotted, and dashed lines, respectively). We used $\dot{J}_{\odot} = 7.169 \times 10^{30} ~ \mathrm{g.cm^2.s^{-2}}$, and $\Omega_{\odot} = 2.87 \times 10^{-6} ~ \mathrm{s^{-1}}$ \citep{WD67}. }
\label{djdt2}
\end{figure*}

With these assumptions, Fig. \ref{djdt2} shows the evolution of $B_*f_*$, $\dot{M}_{wind}$, and $\dot{J}$ as a function of the angular velocity. All three quantities exhibit different regimes, with a steeper dependence on velocity at low spin rates and a shallower slope at high velocities.  Indeed, this reflects the saturation of $f_*$ at high velocities (see Eq.~\ref{bf}). At a given velocity, lower mass stars exhibit higher mean magnetic fields, which primarily results from their larger photospheric pressure. The magnetic field also appears to saturate faster in lower mass stars. This  is a consequence of low-mass stars having longer turnover timescale, while saturation occurs at the same Rossby number for all masses ($Ro_{sat}\simeq 0.1$, cf. Eq.~\ref{bf} and the inset in Fig.~\ref{djdt2}'s upper panel). Unlike previous studies that followed \citet{Kawaler88}'s prescription and used distinct relationships to account for the non-saturated and saturated magnetic regimes, we use a single prescription that smoothly bridges the two regimes. Moreover, as this prescription scales on the Rossby number, it can be straightforwardly extrapolated to any mass range. 

Like the magnetic field, and for the exact same reason, the mass-loss rate first increases with angular velocity at low spin rates and then saturates at higher velocities. The strong mass dependence of the mass-loss rate seen in Fig.~\ref{djdt2} stems from $F_{A*}$ being very sensitive to mass. $F_{A*}$ is a function of $T_{eff}$ and $\log g$ only and rapidly decreases towards lower masses. The angular momentum loss rate can thus be computed at any age step for any mass by combining Eqs.~\ref{djdt}, \ref{ranew}, \ref{bf}, and \ref{mdot}, together with the structural evolution models of \citet{Baraffe98}. Figure~\ref{djdt2} shows the run of angular momentum loss rate as a function of angular velocity. The same trend for a steep increase with rotation at low spin rates followed by a shallower rotational scaling at higher velocities is seen, which naturally results from the combined rotation dependencies of the magnetic field and mass-loss rate discussed above. The asymptotic expressions of the braking law are provided in \citet{Gallet13}. Quantitatively, however, we stress that we had to renormalise the $K_1$ constant appearing in Eq.~\ref{ranew} in a mass-dependent manner to reproduce the observations (see Section \ref{paramK1disc}). With this renormalization, shown in Fig.~\ref{djdt2}, the {braking torque} is found to be roughly the same for all masses in the non-saturated regime, but saturates faster and at lower values for lower mass stars. We emphasise that this behaviour of the {braking torque} indeed holds the keys of the rotational evolution of aging low-mass stars: it encodes both the longer spin down timescale of lower mass stars (due to the weaker {braking torque} in the saturated regime) and their lower final velocities (due to a higher ratio of {braking torque} to stellar moment of inertia in the saturated regime).  We return to this crucial point in Section \ref{paramK1disc}.  

\subsection{Star-disk interaction}
\label{stardisk}

As apparent in Fig.~\ref{compamas},  observations suggest that during the first few Myr of the PMS phase a physical process acts to prevent the stellar surface from spinning up. This behaviour is believed to result from the magnetic interaction between the young stellar object and its accretion disk, even though the actual physical process is not totally elucidated yet \citep[see][]{MP05b,MP08a,MP08b,Zanni11,Zanni2012,PPVI}.

As in \citet{Gallet13} the model developed here assumes that the angular velocity of the stellar surface is held constant as long as the star accretes from its circumstellar disk. The disk lifetime $\tau_{disk}$, a free parameter of the model, thus dictates the duration over which the star is kept at its initial spin rate. When the disk eventually dissipates, after a few Myr, the disk coupling ends, and from thereon, angular momentum is removed at the stellar surface by stellar winds. 

\subsection{Core-envelope decoupling}

The transport of angular momentum in the stellar interior is one of the most important physical processes that occurs during the early MS phase. \citet{Gallet13} showed that this process is central to the rotational evolution of solar-mass stars as it leads to the storage of a large fraction of the stellar angular moment within the rapidly spinning radiative core. As many uncertainties remain in the actual physical process that control angular momentum transport in stellar interiors \citep{Charbonnel05,Charbonnel13}, we adopt here a two-zone model where the inner radiative core and the outer convective envelope, each in uniform rotation, are loosely coupled. We follow \citet{McGB91} in assuming that a quantity $\Delta J$ is exchanged between the core and the envelope over a timescale $\tau_{c-e}$ (hereafter the core-envelope coupling timescale). This quantity $\Delta J$ is the amount of angular momentum that the core and the envelope have to exchange to restore uniform rotation throughout the star, and is given by 
\begin{eqnarray}
\Delta J = \frac{I_{env}J_{core}-I_{core}J_{env}}{I_{core}+I_{env}},
\end{eqnarray}
where $I$ and $J$ refer to the moment of inertia and angular momentum of the radiative core and the convective envelope, respectively. We also assume, as in \citet{Allain98}, that $\tau_{c-e}$ is constant for a given model \citep[see Sect. 3 of][for a more detailed discussion about $\tau_{c-e}$]{Gallet13}.

\subsection{Equation of evolution}

The angular momentum evolution of low-mass stars is controlled by the internal and external torques applied on and within the stars (see Sect. \ref{modcons}). This evolution can be expressed as
\begin{eqnarray}
\frac{dJ}{dt} = \frac{dI_{*}}{dt} \Omega_{*} + I_{*} \frac{d\Omega_{*}}{dt} = \Gamma,
\label{jevol}
\end{eqnarray}
where $J$ is the angular momentum, $I$ the moment of inertia, $\Omega_*$ the angular velocity, and $\Gamma$ the sum of all the external torques. In Eq. \ref{jevol} the quantity $\dot{I}\Omega_*$ is the contraction torque. In the framework of the assumptions described above, the equations that control the angular velocity evolution of the envelope can be expressed using Eq. \ref{jevol} as 

\indent if $t \leq \tau_{disk}$:
\begin{eqnarray}
\Omega_{conv} = \Omega_{init} \nonumber
\end{eqnarray}

if $t > \tau_{disk}$:
\begin{eqnarray}
\frac{d\Omega_{conv}}{dt} = \frac{dJ_{conv}}{dt}\frac{1}{I_{conv}} - \frac{dI_{conv}}{dt}\frac{\Omega_{conv}}{I_{conv}},
\end{eqnarray}
where 
\begin{eqnarray}
\frac{dJ_{conv}}{dt} = -\Gamma_{wind} + \Gamma_{ce} -\Gamma_{rad},
\end{eqnarray}
with $\Gamma_{wind}$ the wind braking torque (see Eq. \ref{djdt}),  $\Gamma_{ce}= \Delta J / \tau_{c-e}$ the quantity of angular momentum transported from the core to the envelope over the timescale $\tau_{c-e}$, and $\Gamma_{rad}$ the rate of angular momentum lost from the convective envelope during the formation of the radiative core. As the core develops, a fraction of the convective envelope becomes radiative and the rate of angular momentum transferred from the envelope to the expanding radiative core is
 \begin{eqnarray}
\Gamma_{rad}=\frac{2}{3} R_{rad}^2 \Omega_{conv} \frac{dM_{rad}}{dt},
\end{eqnarray}
where $dM_{rad}$ is the quantity of material that is contained in a thin shell at a radius $R_{rad}$ inside the star \citep{Allain98}. Finally, the total torque applied on the stellar surface is
\begin{eqnarray}
\frac{dJ_{conv}}{dt} = -\Gamma_{wind}+ \frac{\Delta J}{\tau_{c-e} }  - \frac{2}{3} R_{rad}^2 \Omega_{conv} \frac{dM_{rad}}{dt},
\end{eqnarray}
and the associated angular velocity evolution is
\begin{eqnarray}
\frac{d\Omega_{conv}}{dt} = \frac{1}{I_{conv}}\frac{\Delta J}{\tau_{c-e} } - \frac{2}{3} \frac{R_{rad}^2}{I_{conv}} \Omega_{conv} \frac{dM_{rad}}{dt}  \nonumber \\
 - \frac{dI_{conv}}{dt} \frac{\Omega_{conv}}{I_{conv}} - \frac{\Gamma_{wind}}{I_{conv}}.
\end{eqnarray}
Similarly, the angular velocity evolution of the core is computed as:  
\begin{eqnarray}
\frac{d\Omega_{rad}}{dt} = -\frac{1}{I_{rad}}\frac{\Delta J}{\tau_{c-e} } + \frac{2}{3} \frac{R_{rad}^2}{I_{rad}} \Omega_{conv} \frac{dM_{rad}}{dt}  \nonumber \\
 - \frac{dI_{rad}}{dt} \frac{\Omega_{rad}}{I_{rad}}.
\end{eqnarray}

\section{Results: confronting models to observations}
\label{res}

The free parameters of the model are adjusted so as to best reproduce the observations. These are: the initial rotation period at 1 Myr, $P_{init}$, the core-envelope coupling timescale, $\tau_{c-e}$, the disk lifetime, $\tau_{disk}$, and the calibration constant of the wind braking law, $K_1$. The best parameters found for the slow, median, and fast rotator models are listed in Table \ref{param}.

\begin{table}
\caption{Model parameters.} 
\begin{center}
$M_{*} = 1 ~M_{\odot}$ \\
\label{param}          
\begin{tabular}{c c c c}     
\hline\hline              
Parameter & Slow & Median & Fast \\  
\hline
$P_{init}$ (days) & 8 & 5 & 1.4 \\
$\tau_{c-e}$ (Myr) & 30 & 28 & 10 \\
$\tau_{disk}$ (Myr) & 9 & 6 & 2 \\
$K_1$ & 1.7 & 1.7 & 1.7 \\
\hline                                  
\end{tabular}
\end{center}

\begin{center}
$M_{*} = 0.8 ~M_{\odot}$ \\
\begin{tabular}{c c c c}     
\hline\hline              
Parameter & Slow & Median & Fast \\  
\hline
$P_{init}$ (days) & 9 & 6 & 1.4 \\
$\tau_{c-e}$ (Myr) & 80 & 80 & 15 \\
$\tau_{disk}$ (Myr) & 7 & 5 & 3 \\
$K_1$ & 3 & 3 & 3 \\
\hline                                  
\end{tabular}
\end{center}

\begin{center}
$M_{*} = 0.5 ~M_{\odot}$ \\         
\begin{tabular}{c c c c}     
\hline\hline              
Parameter & Slow & Median & Fast \\  
\hline
$P_{init}$ (days) & 8 & 4.5 & 1.2 \\
$\tau_{c-e}$ (Myr) & 500 & 300 & 150 \\
$\tau_{disk}$ (Myr) & 6 & 3.5 & 2.5 \\
$K_1$ & 8.5 & 8.5 & 8.5 \\
\hline                                  
\end{tabular}
\end{center}

\end{table}

\subsection{The rotational evolution of low-mass stars}

\label{rotevollowmass}

\begin{figure*}[ht!]
     \begin{center}
        \subfigure[$1 ~M_{\odot}$]{%
            \label{model1}
            \includegraphics[angle=-90,width=0.4\textwidth]{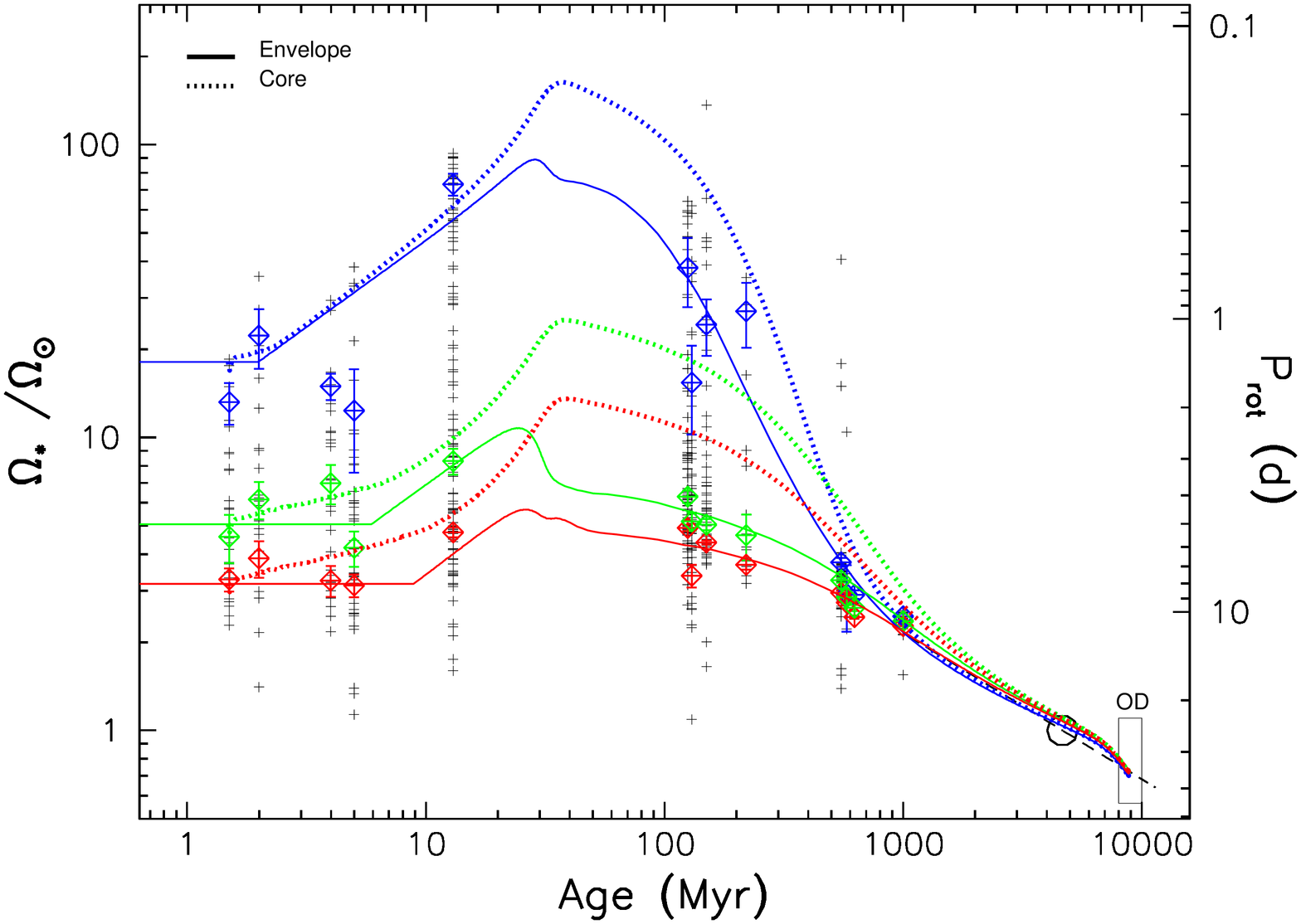}
        }%
        \subfigure[$0.8 ~M_{\odot}$]{%
            \label{model08}
            \includegraphics[angle=-90,width=0.4\textwidth]{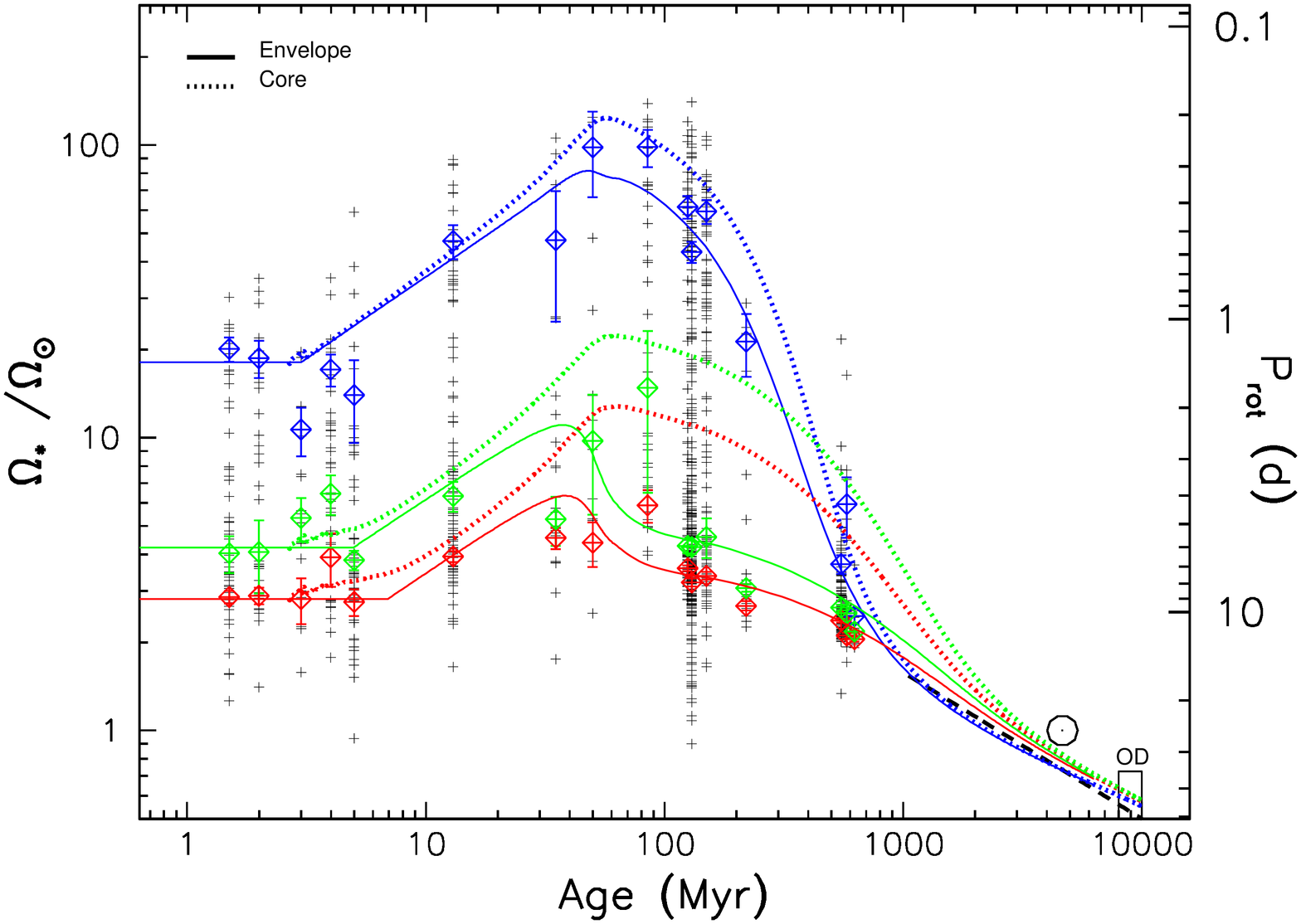}
        }\\
        \subfigure[$0.5 ~M_{\odot}$]{%
           \label{model05}
           \includegraphics[angle=-90,width=0.4\textwidth]{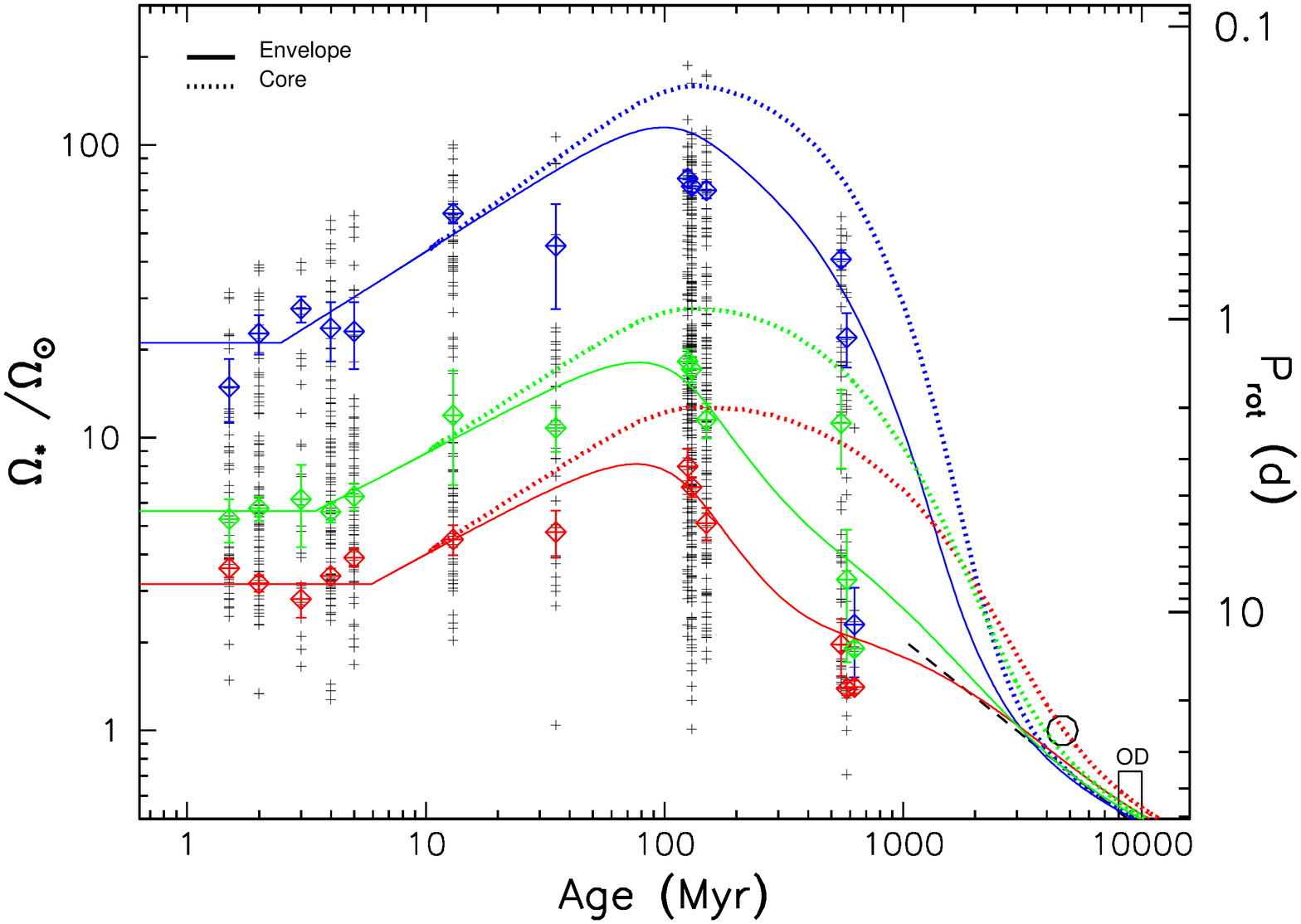}
        }
    \end{center}
    \caption{%
 The angular velocity of the convective envelope (\textit {solid lines}) and of the radiative core (\textit{dashed lines}) is shown as a function of time between 1 Myr and 10 Gyr for slow (\textit{red}), median (\textit{green}), and fast (\textit{blue}) rotator models in three mass bins centred on 0.5 M$_\odot$ (\textit{bottom panel}), 0.8 M$_\odot$ (\textit{upper right panel}) and 1~$M_{\odot}$ (\textit{upper left panel}).  The left vertical axis is labelled with angular velocity normalised to the Sun's, while the right vertical axis is labelled with rotational periods (\textit{days}). The red, green, and blue tilted squares and associated error bars represent the 25$^{th}$, 50$^{th}$, and 90$^{th}$ percentiles of the observed rotational distributions at each sampled age. The black rectangle labelled OD (\textit{lower right corner of each panel}) shows the angular velocity dispersion of old disk field stars. The open circle is the angular velocity of the present Sun shown for reference, and the dashed black line illustrates Skumanich's (1972) relationship, $\Omega\propto t^{-1/2}$.  }%
   \label{model}
\end{figure*}

Figure \ref{model} summarises the angular velocity evolution of slow, median, and fast rotators in the three selected mass bins, centred on 0.5, 0.8, and 1 M$_\odot$.

\begin{figure*}[ht!]
     \begin{center}
        \subfigure[$1 ~M_{\odot}$]{%
            \label{sddww1}
            \includegraphics[angle=-90,width=0.4\textwidth]{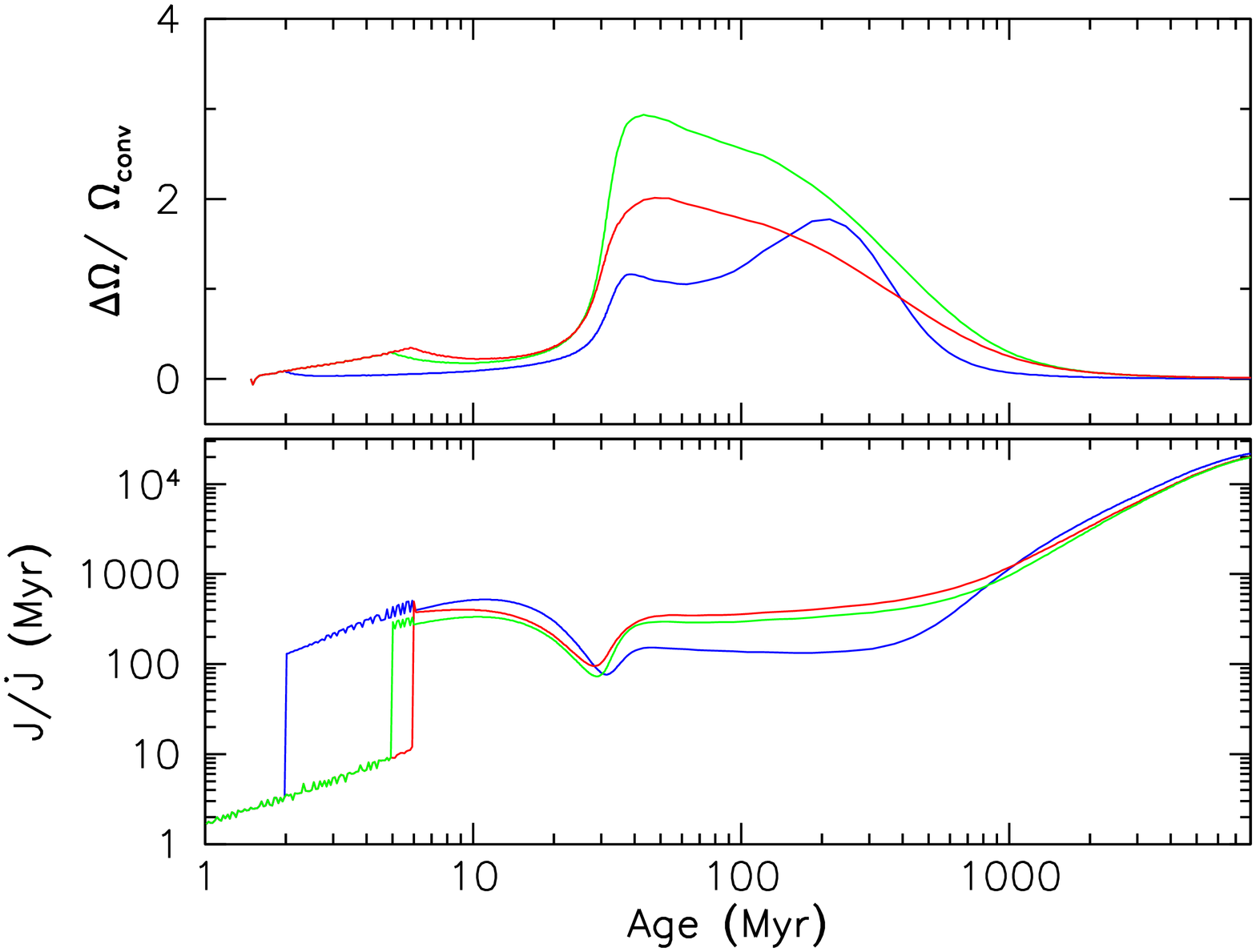}
        }
        \subfigure[$0.8 ~M_{\odot}$]{%
           \label{sddww08}
           \includegraphics[angle=-90,width=0.4\textwidth]{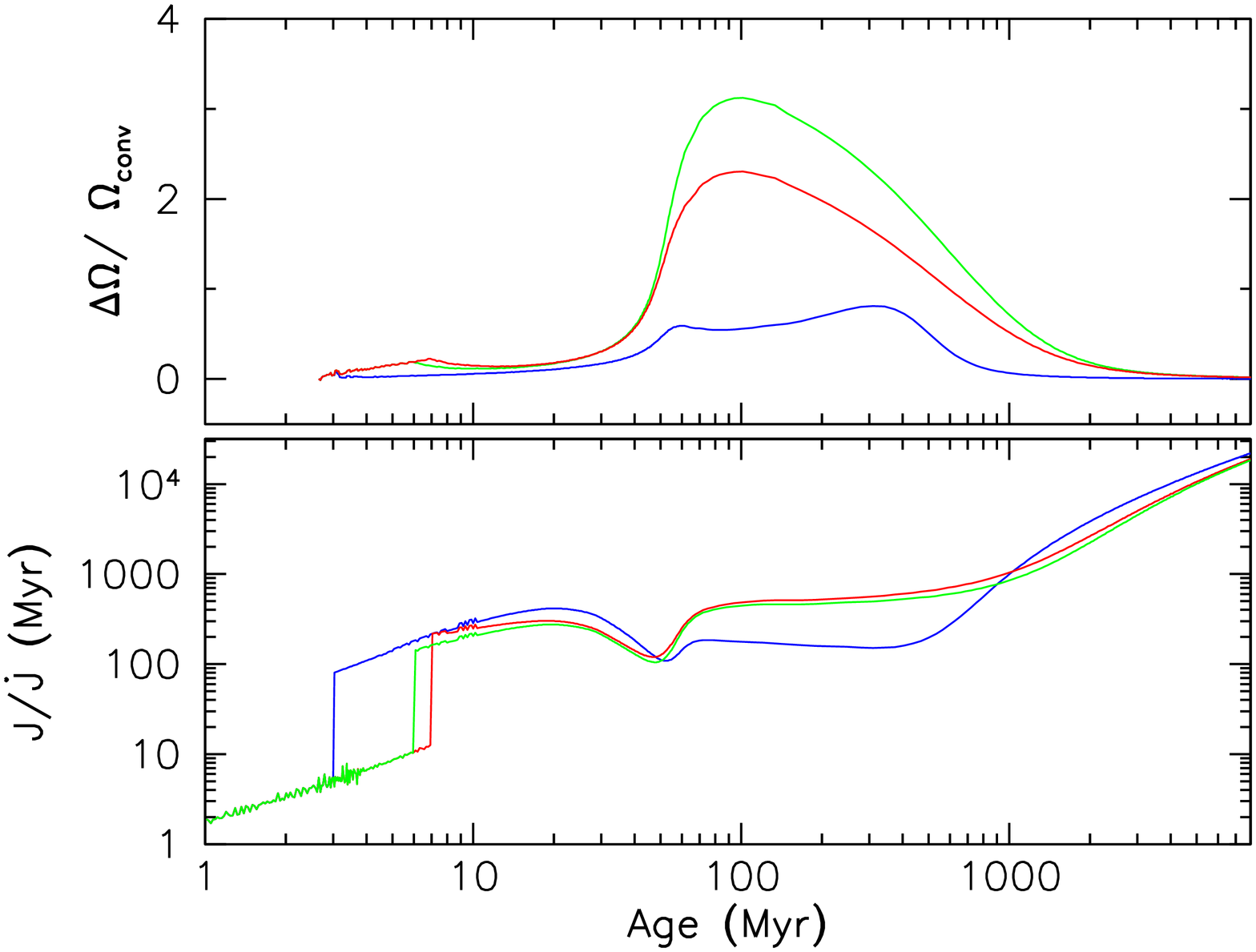}
        }\\ 
        \subfigure[$0.5 ~M_{\odot}$]{%
            \label{sddww05}
            \includegraphics[angle=-90,width=0.4\textwidth]{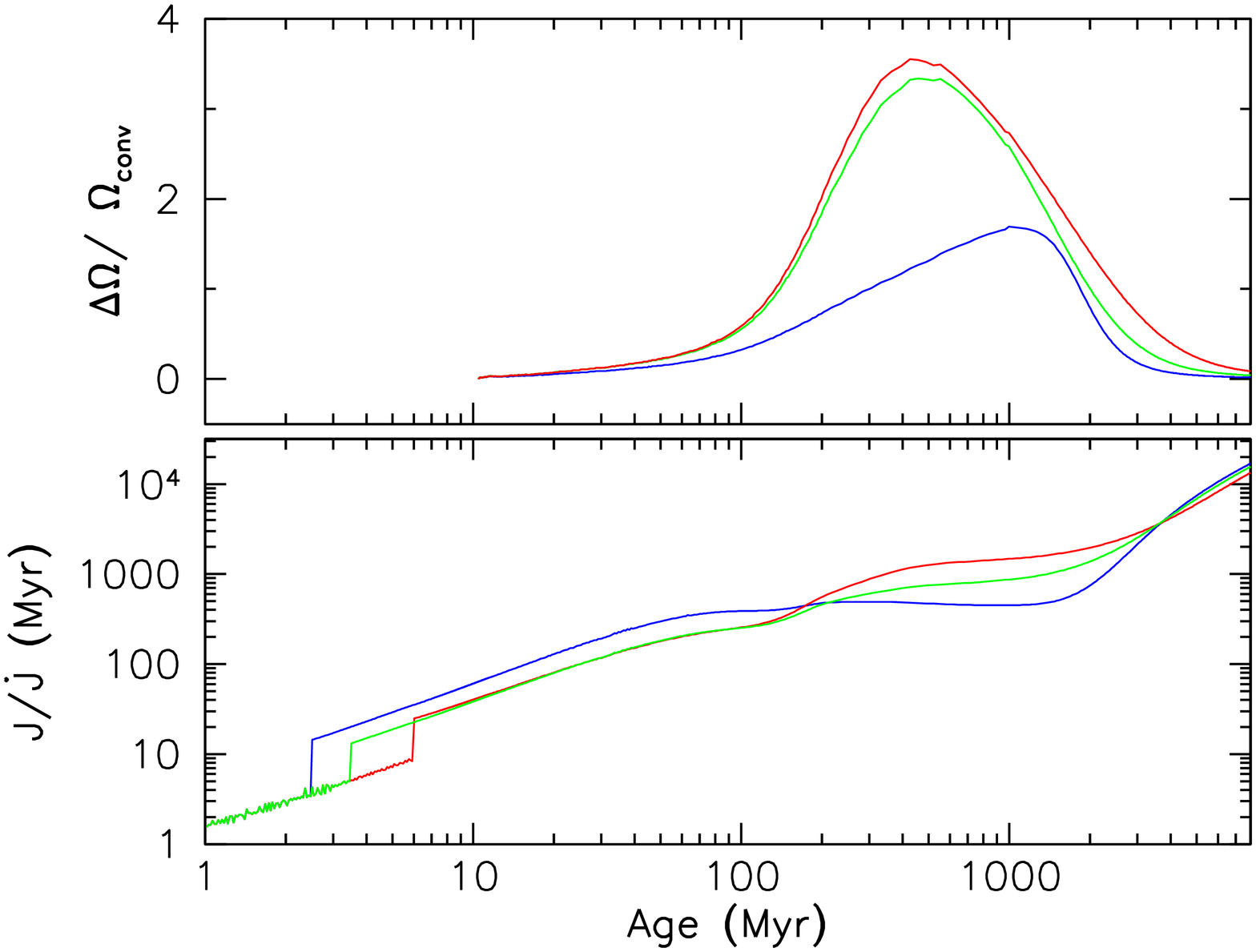}
        }%
    \end{center}
    \caption{%
\textit{Upper panel:} Velocity shear at the base of the convective zone $(\Omega_{core}-\Omega_{env})/\Omega_{env}$ in the case of fast (blue), median (green), and slow (red) rotator models. \textit{Lower panel:} Spin-down timescale ($|  J/\dot{J} |$) expressed in Myr, in the case of fast (blue), median (green), and slow (red) rotator models. \textit{a) $1 ~M_{\odot}$}, \textit{b) $0.8 ~M_{\odot}$}, and \textit{c) $0.5 ~M_{\odot}$}.
     }%
   \label{sddww}
\end{figure*}

\subsubsection{Solar-mass stars} 

Solar-mass models were already presented in \citet{Gallet13}. The data set has been updated for this study by adding the results recently published for five additional open clusters, namely: Cep OB3b (4 Myr), IC2391 (50 Myr), $\alpha$ Per (85 Myr), NGC2516 (150 Myr), and M34 (220 Myr), although not all of them have enough stars in a given mass bin to constrain the models (see Table~\ref{opencluster}).  We also restricted the mass bin to stars strictly within the 0.9-1.1 M$_{\odot}$ mass range. As a result, the values of the model parameters have changed slightly, and are listed in Table~\ref{param}. Compared to the values listed in \citet{Gallet13}, we fixed the $K_1$ parameter to 1.7 for the three rotator models (instead of 1.8 for the slow and median rotators, and 1.7 for the fast rotators), and used a longer disk's lifetime for the slow rotator model (9 Myr instead of 5 Myr). Other parameters did not change by more than 20\% from the previous study. These updates do not significantly affect the overall angular velocity evolution of solar-like stars and the results and discussion provided in \citet{Gallet13} remain valid.

\subsubsection{Stars of 0.8 M$_\odot$}

Figure~\ref{model08} and \ref{model1} show a qualitatively similar angular momentum evolution for 0.8 and 1.0 M$_\odot$ stars. The initial rotation rates, 1.4, 6, and 9 days for rapid, median, and slow rotators, respectively, are fixed by the percentiles of the rotational distributions of the youngest clusters. The rapid spin up between the early PMS and the 13 Myr h Per cluster requires a disk lifetime as short as 3~Myr for rapid rotators, increasing to 7 Myr for slow rotators. To rapidly accelerate the convective envelope up to the ZAMS where $\Omega_*^{ZAMS}\simeq 50-60~\Omega_\odot$, rapid rotators must maintain nearly solid-body rotation, which requires a core-envelope coupling timescale as short as 15 Myr. In contrast, median and slow rotators have moderate velocities on the ZAMS, which calls for a longer core-envelope coupling timescale of 80 Myr, so that the envelope is readily braked while the core still spins fast. In these models, the rotational gradient between the core and the envelope reaches a maximum of $\Delta\Omega/\Omega\simeq 2-3$ at about 100 Myr (cf. Fig.~\ref{sddww08}). 

As in the case of a 1 $M_{\odot}$ model, we find that the core-envelope coupling timescale is similar in slow and moderate rotators, and much longer than for fast rotators. In turn, such a long coupling timescale implies that the resurfacing of the angular momentum stored in the radiative core will occur over an extended period. Indeed, this long-term outward transport of angular momentum partly compensates for the angular momentum losses at the stellar surface due to stellar winds. This results in a relatively shallow velocity decrease on the early main sequence, as observed for slow and moderate rotators that exhibit a clear change in the spin-down slope between 100 and 500 Myr (cf. Fig~\ref{model08}). In contrast, rapid rotators are spun down much more {easily} on the early MS, as the {braking torque} scales with surface velocity.  

The angular velocity of the slow, median, and fast rotators eventually converge towards the same spin rate at about 1 Gyr. By the age of the Sun, all the models presented here exhibit a nearly solid body rotation and have a surface velocity lower than the Sun's, in agreement with the Kepler results \citep[see][]{McQuillan2013}. The lower terminal velocity of 0.8 M$_\odot$ models compared to the solar-mass models directly stems from the calibration of the braking law. As shown in Table~\ref{param}, the $K_1$ constant of the braking law was increased from 1.7 for solar-mass stars to 3 for 0.8 M$_\odot$ stars. Indeed, regardless of the detailed rotational history, the terminal velocity on the main sequence only depends upon the normalization of the braking law, thus decreasing as the $K_1$ constant increases. 
Note that with this renormalization, the {braking torque} of lower mass stars in the saturated regime remains weaker than for solar-mass stars (cf. Fig.~\ref{djdt2}), which simultaneously accounts for their longer spin-down timescale on the main sequence. 

\subsubsection{Stars of 0.5 M$_\odot$}

The main trends of the rotational evolution of solar-mass stars can still be recognised in that of 0.5~M$_\odot$ stars, as shown in Fig.5. A clear difference, however, is that the evolutionary timescales become much longer at lower masses. In particular, the rotational spin down occurs over several Gyr on the main sequence for 0.5~M$_\odot$ stars, compared to only a few 100 Myr for solar-mass stars. 

The early PMS evolution is not very different from that of more massive stars: similar disk lifetimes, increasing from 2.5 Myr for fast rotators to 6 Myr for slow rotators, are required to prevent the lower mass stars from spinning up.  Once the disks are dissipated, the spin up occurs on a longer timescale to the ZAMS, which is reached at about 120 Myr. As for higher mass stars, even though stellar winds are active during this phase, they are unable to balance the acceleration torque due to decreasing stellar moment of inertia. 

In comparison to the 1 $M_{\odot}$, the core-envelope coupling timescale increases by a factor of 10 or more in 0.5~M$_\odot$ models, amounting to 150 Myr for fast rotators and 500 Myr for slow and moderate rotators. One recovers the increasing coupling timescale from slow to fast rotators, as was the case for 0.8 and 1~M$_\odot$ models. This allows initially fast rotators to reach high velocities at the ZAMS, while the convective envelope of slow and moderate rotators is already significantly braked before the star reaches the ZAMS (see Fig.5). Indeed, the increased core-envelope decoupling in slower rotators, coupled to their systematically longer disk lifetime, is at the origin of the enhanced dispersion of rotational velocities on the ZAMS. 

In these models, the largest amount of differential rotation between the inner radiative core and the outer convective envelope is reached at {500 Myr} for slow rotators and amounts to $\Delta\Omega/\Omega\simeq 3$ (cf. Fig.~\ref{sddww05}).  The resurfacing of angular momentum being transported from the core to the envelope on a timescale of $\leq$1 Gyr on the MS is still seen in Fig.~\ref{model05} as a kink in the spin-down rate of slow rotators, even though it is not as pronounced as in higher mass stars.  This is also highlighted in Fig. \ref{sddww05} (\textit{lower panel}) where a small plateau can be seen at $\approx$ 400 Myr in the spin-down timescale of slow rotators. 

The angular velocity of the slow, median, and fast rotators models eventually converges towards the same rotation rate at about 3 Gyr in a state of nearly uniform rotation (see Fig.~\ref{model05} and \ref{sddww05}). As explained above for 0.8~M$_\odot$ stars, the rotational convergence is delayed for lower mass stars due to the weaker {braking torque} in the saturated regime (cf. Fig.~\ref{djdt2}). However, when rotational convergence is completed, lower mass stars exhibit lower terminal velocities, due to their reduced moment of inertia. Indeed, at low velocities the ratio of {braking torque} to moment of inertia increases for lower mass stars (cf. Fig.~\ref{djdt2}), i.e. lower mass stars are more effectively braked in the non-saturated regime, which directly results in lower terminal velocities.  

\subsection{Differential rotation}

The radial differential rotation rate is defined here by $\Delta \Omega / \Omega_{conv}$ where $\Delta \Omega = \Omega_{core}-\Omega_{env}$. This quantity measures the velocity shear at the tachocline between the radiative core and the convective zone of the two-zone model. This is obviously a simplification of the actual rotation profile expected to arise in the stellar interior from angular momentum redistribution. Nevertheless, it provides a measure of how much angular momentum may be stored in the stellar interior at any given time of its evolution. 

Figure \ref{sddww} shows the evolution of differential rotation for the models described above. A few systematic trends emerge. Somewhat unexpectedly, slow and moderate rotators have systematically larger differential rotation rates than fast rotators. This is the direct result of the assumption of longer core-envelope coupling timescale for slower rotators (cf. Table~\ref{param}). This assumption is required to account for the widening of the rotational velocity dispersion from the early PMS to the ZAMS: initially fast rotators have to be spun up quite {effectively} while the surface of initially slow rotators must be spun down before they reach the ZAMS. The only way to achieve {this, supposing realistic disk lifetimes ($\leq$ 10 Myr),} is to have a weaker core-envelope coupling in slow rotators than in fast rotators. The direct consequence of this assumption is the stronger differential rotation in slow and moderate rotators on the ZAMS, with $\Delta \Omega / \Omega_{conv}\simeq 2-3$ compared to $\leq$1 for fast rotators. 

Another clear trend is that the peak of differential rotation rate occurs earlier in slow and moderate rotators than in fast rotators. While the former exhibit a maximum tachocline shear at or close to the ZAMS, the maximum occurs later on the main sequence for the latter (cf. Fig.6). This differential effect presumably results from the complex interplay between the spin-down rate, which is steeper for fast rotators, and the timescale for angular momentum distribution, which is shorter for fast rotators (shorter core-envelope coupling timescales, cf. Table~\ref{param}).  Clearly, the evolution of differential rotation in stellar interiors is relatively complex and depends on the details of the models. 

Overall, the two-zone models suggest that a large quantity of angular momentum can be hidden in the stellar interior over duration of a few 100 Myr and up to about 1 Gyr on the main sequence. This has obvious consequences on stellar evolution \citep{Charbonnel13}, the depletion of light elements such as Lithium \citep{Bouvier08,Eggenberger12}, and the type of magnetic dynamos that may be instrumental in young solar-type stars \citep{Vidotto2014}.  We return to these aspects in the next section.

\section{Discussion}
\label{disc}

Thanks to the measurement of thousands of rotation periods in star-forming regions and young open clusters, we are now able to accurately trace the angular momentum evolution of low-mass stars from the birthline ($\leq$ 1 Myr) to the age of the Sun. Over the mass range considered here, three main phases of rotational evolution can be identified: a nearly constant surface rotation rate during the first few million years of the PMS phase, a sharp increase towards the ZAMS, and a steady decline on the early MS on a timescale of a few 100 Myr. Furthermore, observations suggest a large initial spread of rotation periods for the youngest star-forming regions. At the start of the PMS phase, rotational distributions exhibit periods ranging from 1-3 days to 8-10 days. This initial spread increases further at ZAMS, with periods ranging from 0.2-0.4 days to about 6-8 days. Eventually, the spread is erased on the early MS over a timescale of a few 100 Myr, as surface rotation converges to a low, mass-dependent terminal velocity. The models developed in the previous section show that these trends can be well reproduced with a small number of assumptions: i) a magnetic star-disk interaction that is believed to prevent the stars from spinning up during the early PMS, which results in a nearly constant rotation rate during the disk accretion phase; ii) angular momentum losses due to magnetised stellar winds, a process whose magnitude depends on stellar mass, magnetic field, and rotation rate, and which dominates the rotational evolution of low-mass stars past the ZAMS; and iii)  angular momentum redistribution in the stellar interior, which allows a large fraction of the initial angular momentum to be temporarily stored in the inner radiative core on the early MS. 

The combination of star-disk interaction, wind braking, and core-envelope decoupling fully dictates the surface evolution of low-mass stars. In the following sections, we discuss the relevance of each model parameters and their dependence on mass.

\subsection{Initial conditions and disk lifetimes}

In the early PMS, the central model parameter is the disk lifetime, $\tau_{disk}$. Since a constant rotation rate is assumed during the accretion phase, the longer the disk lifetime, the lower the rotation rate at ZAMS. Thus, disk lifetimes of a few Myr are required by the models to reproduce the wide rotational distributions observed at ZAMS. Recent measurements of infrared excesses and disk accretion rates indicate that most stars are born with a disk, while only half of them still have an accreting disk at about 3 Myr, and very few indeed by an age of 10 Myr  \citep[e.g.][]{Hernandez08,Wyatt08,Williams11}. Observations thus suggest an average disk lifetime of 3-5 Myr, and a maximum lifetime of order of 10-20 Myr  \citep{Bell13}. This is quite consistent with the disk lifetimes required by the models, which range from 2 to 9 Myr (cf. Table~\ref{param}). 

The combination of the initial period, $P_{init}$, and the disk lifetime, $\tau_{disk}$, primarily drives the surface velocity at ZAMS. There is some degeneracy in the models between these two parameters, as a longer initial period could be compensated by a shorter disk lifetime in order to yield the same velocity at ZAMS. To solve this degeneracy, we fixed the initial period based on the observed rotational distributions of the youngest PMS clusters. As a result, we find that disk lifetimes do not vary much with stellar mass but are systematically longer for slow rotators than for fast rotators in each mass bin (cf. Table~\ref{param}). As discussed in \citet{Gallet13} for solar-mass stars, this correlation between disk lifetime and initial period, which is shown here  to extend to lower mass stars as well, may be a sequel to the star-disk interaction process operating in the embedded phase. In other words, more massive disks would yield lower initial velocities and live longer. Hence, a distribution of protostellar disk masses might actually be the source of the dispersion of initial rotation rates of low-mass stars at the start of their PMS evolution.

\subsection{Core-envelope decoupling and the timescale for angular momentum transport}

Past the disk regulation phase, the \textit{shape} of the gyrotracks is mostly dictated by the equilibrium between the wind braking torque and the internal redistribution of the angular momentum. The dominant angular momentum transport process has not been securely identified yet, but several candidates have been proposed, namely hydrodynamical instabilities \citep[see e.g.][]{Kri97}, internal magnetic fields \citep[see e.g.][]{Den07,Den10,Spada10,Eggenberger12}, and internal gravity waves \citep[see][]{Talon05,Talon03,Charbonnel13}. 

Here, we used a simplified two-zone model consisting of a radiative core and a convective envelope, which are both in uniform rotation but at a different rate. The assumption of core-envelope decoupling thus leads to a velocity shear, indeed a discontinuity, at the tachocline. This obviously should be considered as a simplified approximation of more complex internal rotational profiles \citep[e.g.][]{Spada10,Den10,Brun11,TC2011,Lagarde12}. As described in Sect. \ref{modcons}, we consider that the angular momentum transport mechanism acts to restore uniform rotation between the core and the envelope. The central parameter is the core-envelope coupling timescale, $\tau_{c-e}$, which is the time it takes to reach a state of uniform rotation. The efficiency of the angular momentum transport process is then measured by $\tau_{c-e}^{-1}$. 

A striking result of our models is that $\tau_{c-e}$ appears to strongly depend both on mass and on spin rate. As Table~\ref{param} summarises, lower mass stars have coupling timescales at least 10 times longer than solar-mass stars, and slow rotators have coupling timescales 3 to 6 times longer than fast rotators.  Indeed, the varying coupling timescales are mostly responsible for producing the noticeable differences exhibited by the slow and fast rotator models presented in the previous sections. 

A small value of $\tau_{c-e}$ for fast rotators will lead to high rotation rate at ZAMS since angular momentum is {effectively} transported from the core to the envelope over a timescale shorter than the contraction timescale. Once on ZAMS, rapid spin down will occur because of the $\Omega$-dependence of the wind braking law. This accounts for the triangular shape of the fast gyrotracks relative to the ZAMS (cf. Fig.5). In contrast, the much longer  $\tau_{c-e}$ of slow rotators results in a more complicated gyrotrack shape, which peaks slightly before the ZAMS as the envelope is spun down first and flattens out on the early MS as angular momentum hidden in the core resurfaces in the envelope over several  $\tau_{c-e}$.  The main differences between slow/moderate and fast rotator models thus primarily arise from their widely different coupling timescales. It is worth noticing that this behaviour is true over the full mass range investigated here. 

\citet{Bouvier08} and \citet{Irwin09a} developed similar models for solar-mass stars. They find a coupling timescale of 110 Myr for slow rotators and 6 Myr for fast rotators. In comparison, our models yield 30 and 10 Myr, respectively. While the values for slow rotators broadly agree, the coupling timescale for fast rotators differs widely between our models. A major difference between former and current models lies in the adopted wind braking law. While \citet{Bouvier08} and \citet{Irwin09a} used \citet{Kawaler88} braking law, we use here \citet{Matt12} prescription. As the latter braking law is more {effective} at extracting angular momentum than the former \citep[cf. Fig.8 of ][]{Gallet13}, a shorter core-envelope coupling timescale compensates for the enhanced surface {braking torque} and produces similar spin-down timescales on the MS. This instance clearly illustrates how the absolute values of $\tau_{c-e}$ depend on the assumed wind {braking torque}.

\begin{figure}
\centering
\includegraphics[angle=-90,width=7cm]{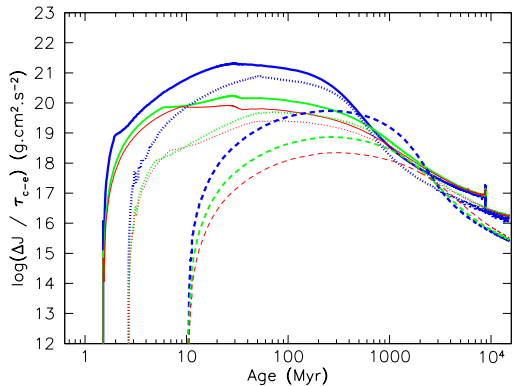}
\caption{The angular momentum transport rate between the radiative core and the convective envelope is shown as a function of time for the three rotator models: slow (red), median (green), and fast (blue), and for the three mass bins explored here: 1 $M_{\odot}$ (solid), 0.8 $M_{\odot}$ (dotted), and 0.5 $M_{\odot}$ (dashed).}
\label{deltatau}
\end{figure}

The coupling timescale $\tau_{c-e}$ we derived from our model also significantly increases towards lower masses, a result that is consistent with those of \citet{Irwin07} and \citet{Den10}. Figure \ref{deltatau} displays the evolution of the transport rate $\Delta J/ \tau_{c-e}$ as a function of time for the three rotator models over the three mass bins considered here.
The angular momentum transport rate of a 0.5 $M_{\odot}$ star is 2 orders of magnitudes smaller than for a 1 $M_{\odot}$ star, as the coupling timescale shortens from $\approx$ 300 Myr to $\approx$ 30 Myr for 0.5 and 1 $M_{\odot}$, respectively.

Another way to estimate the angular momentum exchange between the core and the envelope is to estimate the viscosity $\nu$ involved in the angular momentum transport mechanism. \citet{Den10} converted this viscosity to a characteristic timescale that can be compared to the core-envelope coupling timescales  used here. This conversion can be expressed as $\tau_{c-e} = 6.3\times10^{20}~\nu^{-1.208}$, with $\nu$ in cm$^2$s$^{-1}$. The characteristic timescale associated with angular momentum transport in the stellar interior is related to the turbulent viscosity $\nu$ \citep{Heger00} by
\begin{eqnarray}
\tau_{c-e} = l^2 / \nu
\end{eqnarray}
where $l$ is the characteristic length of the redistribution current. Hence, identifying the constant $6.3\times10^{20}$ above to $l^2$, we find a scale of  $l = 2.51 \times 10^{10}$ cm i.e. $40\%~R_*$ for a 1 $M_{\odot}$ star. This suggests that the diffusive approach and the two-zone models are somewhat equivalent and yield similar results. 
\begin{figure}
\centering
\includegraphics[angle=-90,width=9cm]{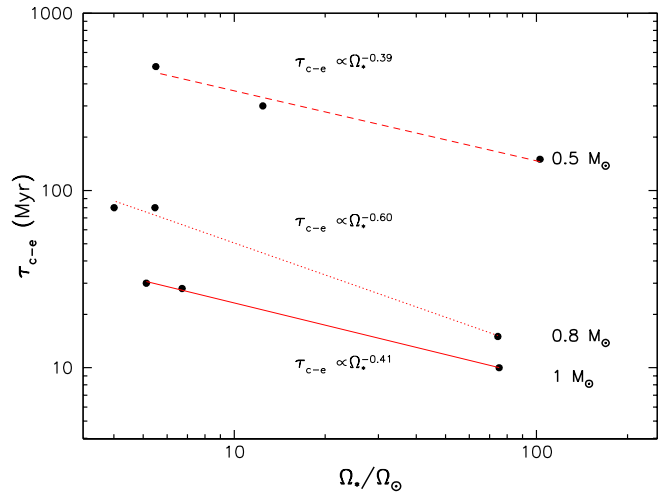}
\caption{Coupling timescale as a function of surface angular velocity at ZAMS for three mass bins: 1 M$_{\odot}$ (solid), 0.8 M$_{\odot}$ (dotted), and 0.5 M$_{\odot}$ (dashed). The values derived from our models are shown as black dots (cf. Table~\ref{param}) and the red lines represent a power-law fit to the data. }
\label{taudecwcz}
\end{figure}

New constraints on the angular momentum transport rate in stellar interiors have emerged from the recent asterosismic results derived from the analysis of Kepler light curves. In particular, the detection of mixed gravity and pressure modes have allowed various groups to derive the internal rotation profile of red giant stars, evolving off the main sequence \citep[e.g.][]{Deheuvels12,Mosser12}. Using a diffusive approach, \citet{Eggenberger12b} showed that an enhanced  "anomalous" viscosity of $\nu$ = 3 $\times$ 10$^4$~cm$^2$s$^{-1}$ is required to account for the rotational splitting of the red giant KIC 8366239, whose rotational profile is much shallower than diffusive models would predict. Using the conversion we derived above from \citet{Den10}, this translates into a coupling timescale of 78 Myr, i.e. not very different than the one we derived from our PMS-MS models. This opens the intriguing possibility that we might be dealing with the same physical process being instrumental from the PMS to the MS and beyond redistributing angular momentum in stellar interiors, thus reducing the amount of rotational shear to be expected at any evolutionary phase. 
 
Ideally, one would hope to derive coupling timescales directly from the theory of angular momentum transport processes. In practice, a characteristic timescale for angular momentum redistribution is difficult to derive from analytical calculations and numerical simulations. In hydrodynamical simulations, such as the STAREVOL model \citep{Siess00}, the angular momentum redistribution timescale can be linked to physical mechanisms such as meridional circulation and thermal diffusivity. For these processes, the coupling timescale varies from few 100 Myr to few Gyr (L. Amard, private communication). For internal gravity waves the associated coupling timescale is about 10 Myr \citep{Zahn97}. 

A recent attempt to derive coupling timescale analytically has been performed by \citet{Oglethorpe13}. Assuming that a large fraction of the radiative core is held in solid-body rotation by the internal magnetic field that is confined below the tachopause (i.e. the base of the tachocline), while angular momentum is transferred through the tachocline by large-scale meridional flows, they derive a coupling timescale proportional to the local Eddington-Sweet timescale across the tachocline finding
\begin{eqnarray}
\tau_{c-e} \simeq t_{ES} \times \frac{I_{core} I_{conv}}{I_{tc}(I_{core}+I_{conv})},
\end{eqnarray}
with 
\begin{eqnarray}
t_{ES} = N_{tc}^2/2\Omega_{conv}^2 \times (\delta/R_{core})^2 \times \delta^2/\kappa_{tc},
\end{eqnarray} 
where $I_{tc}$, $\kappa_{tc}$, $N_{tc}$, and $\delta$ are the moment of inertia, thermal diffusivity, buoyancy frequency, and thickness of the tachocline, respectively.

With these analytic expressions, and using the parameter values of their reference model, the average coupling timescale would be $\bar{\tau}_{c-e}=0.34$ Myr for the 1 M$_{\odot}$ stars. For the 0.8 and 0.5 $M_{\odot}$, we used the values of the density, buoyancy frequency, and thermal diffusivity extracted from the STAREVOL model \citep[][ Louis Amard, private communication]{Siess00} and extrapolated \citet{Oglethorpe13} model to derive $\tau_{c-e}$=8 Myr and $\tau_{c-e}$=60 Myr, for 0.8 and 0.5 M$_{\odot}$, respectively, adopting the surface velocities of median rotators at ZAMS. We caution, however, that we computed these values assuming the same tachocline thickness for lower mass stars as for solar-mass stars, while thicker tachoclines would lead to much longer coupling timescales (cf. Eq. 15). 
Even though these values do not match those we empirically derive from parametric models, we note that theoretically predicted coupling timescales increase towards lower masses as required by the angular momentum evolution models. 

\citet{Oglethorpe13}'s derivation further predicts coupling timescales that are inversely proportional to the square of the surface velocity, i.e. $\tau_{c-e} \propto 1/\Omega_{conv}^2$. Figure \ref{taudecwcz} shows the variation of the coupling timescales upon angular velocity as derived  from our parametric models. A power-law fit yields $\tau_{c-e} \propto 1/\Omega_{conv}^{[0.4 - 0.6]}$, while a previous study by \citet{Moraux13} of PMS and ZAMS clusters suggested $\tau_{c-e} \propto 1/\Omega$. While these parametric slopes are at variance with the analytical predictions, they nevertheless agree with the expected trend for a shorter coupling timescale in faster rotators. 

\subsection{The mass-dependent efficiency of magnetic braking}
\label{paramK1disc}

As shown in Sect. \ref{res}, and discussed in \citet{Gallet13}, the adopted braking law is a crucial parameter of angular momentum evolution models, as it dictates most of the rotational evolution on the main sequence. In the model presented here, the recent results of the numerical simulations from \citet{Matt12} and \citet{Cranmer11}  have been combined to infer the main properties of the surface wind generated by low-mass stars. This new braking law contrasts with the well-known \citet{Kawaler88} prescription used in almost all angular momentum evolution models so far \citep[e.g.][]{Bouvier08,Irwin09a,Den10,Spada11} as well as with the Kawaler-modified model proposed by \citet{Reiners2012}. A detailed comparison between these braking laws is provided in Sect. 5.2 of \citet[][]{Gallet13} \citep[see also][]{Matt2014}. 

In the \citet{Matt12} prescription adopted here, together with the dynamo relationship and mass-loss estimates from \citet{Cranmer11}, the predicted {braking torque} already includes the mass dependency. 
This is seen in Eqs. \ref{djdt}-\ref{mdot}, where mass-dependent parameters appear explicitly (namely: R$_\star$, $v_{esc}$, T$_{eff}$, $Ro$, L$_\star$, $\rho_\star$). Hence the braking law takes the form 
\begin{eqnarray}
\Gamma_{wind} = K_1^2 \times f(\Omega_\star, M_\star, R_\star, B_\star,\dot M_\star,...),
\end{eqnarray}
where $K_1$, the calibration factor, is expected to be a constant for a given magnetic topology. Thus, in \citet{Matt12}, who assumed a large-scale dipolar field, $K_1$=1.30.    

We were unable to fit the observed rotational evolution of low-mass stars with a constant wind braking calibration factor over the mass range explored here. As summarised in Table~\ref{param}, the $K_1$ parameter had to be recalibrated to a value of 1.7, 3.0, and 8.5, for 1 M$_\odot$, 0.8 M$_\odot$, and 0.5 M$_\odot$ models, respectively. The $K_1$ factor was, however, kept the same at a given mass for the slow, median, and fast rotator models. The increasing $K_1$ factor towards decreasing mass suggest that the efficiency of the magnetic braking by stellar winds increases towards lower masses\footnote{Note that the {braking torque} always remains higher for higher mass stars, as shown in Fig.~\ref{djdt2}.}. We offer two possible explanations for this result. 

The stellar wind simulations of \citet{Matt12} computed for a dipolar magnetic geometry were expanded to quadrupolar and octupolar magnetic topologies by \citet{Reville14}. As expected, all other parameters being equal, the radially averaged Alfv\'en radius is found to be systematically smaller for more complex magnetic geometries. Hence, our empirical recalibration of the $K_1$ factor could be interpreted as indicating that higher mass stars have more complex magnetic geometries than lower mass stars, thus reducing the efficiency of magnetic braking by stellar winds at higher masses. A mass-dependent magnetic topology is indeed suggested for main-sequence stars, with solar-mass stars having more multipolar magnetic structure than lower mass stars whose magnetic topology tends to be dipole dominated \citep{Petit08,Donati09,Morin10,Gregory2012}. 

A more prosaic explanation, however, may simply be that the mass-loss rate prescription that we use depends too strongly on stellar mass. As seen from Eq. \ref{djdt} and \ref{ranew} we find
\begin{eqnarray}
 \Gamma_{wind} \propto K_1^2 \times \dot M_{wind}^{1-2m} \simeq K_1^2 \times \dot M_{wind}^{0.56}
 \end{eqnarray}
 so that we could produce the same results by varying the mass-loss rate instead of recalibrating the $K_1$ factor. Keeping the $K_1$ factor to 1.30 as in \citet{Matt12} would require multiplying the mass-loss rates of 1 M$_\odot$, 0.8 M$_\odot$, and 0.5 M$_\odot$ stars by factors of 2.6, 
20, and 860, respectively.  Compared to the mass-loss rates used in the parametric models above (see Fig.~\ref{djdt2}), which range from 10$^{-16.5}$ M$_\odot$yr$^{-1}$ for 0.5 M$_\odot$ stars to 10$^{-13}$ M$_\odot$yr$^{-1}$ for young solar-mass stars, this would translate into mass-loss rates in the range $\sim$10$^{-14}$-10$^{-13}$ M$_\odot$yr$^{-1}$ over the whole mass range. The latter values might actually be in better agreement with the observations \citep[e.g.][]{Vidotto2013}. Clearly, additional measurements of magnetic topologies and mass-loss rates are needed for young solar-type stars to be able to better calibrate the {braking torque} they undergo as they evolve on the early main sequence. 
 
\begin{figure}
\centering
\includegraphics[angle=-90,width=8cm]{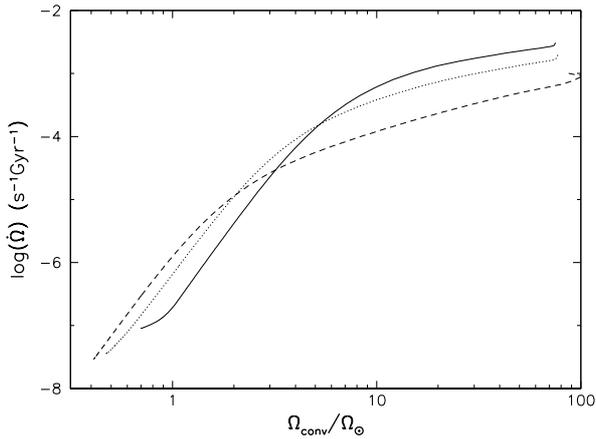}
\caption{The spin-down rate $\dot \Omega=\dot{J}/I_*$ is plotted as a function of angular velocity for 1 M$_{\odot}$ (solid), 0.8 M$_{\odot}$ (dotted), and 0.5 M$_{\odot}$ (dashed) stars. Note the cross-over of the deceleration curves as the stars go from the magnetically saturated to the unsaturated regimes.}
\label{deltajI}
\end{figure}

As discussed in Sect. \ref{rotevollowmass}, lower mass stars take longer to spin-down on the main sequence, but once they have reached rotational convergence, their spin rate is lower than that of solar-mass stars (cf. Fig. \ref{model}). The longer spin-down timescale and lower terminal velocities of lower mass stars both directly results from the braking law. Since stars in the 0.5-1.0 M$_\odot$ mass range have about the same maximum velocities at ZAMS, the spin-down timescale primarily depends on how much angular momentum is extracted in the magnetically saturated regime. As shown in Fig.~\ref{djdt2}, this quantity decreases towards lower mass stars, thus accounting for their longer spin-down timescale. The terminal velocity, however, depends mostly on the extraction of angular momentum once the star is nearing rotational convergence, i.e. in the unsaturated regime (low rotation rate). Fig.~\ref{djdt2} shows that the {braking torque} is then about the same for all stars in the 0.5-1.0 M$_\odot$ mass range, and since the lower mass stars have a lower moment of inertia, they are more {easily} braked, eventually reaching lower terminal velocities. Hence, the two major properties of the rotational evolution of low-mass stars on the main sequence are fully encoded into the mass-dependent braking law.  This is illustrated in Fig.~\ref{deltajI}, which shows the deceleration rate of low-mass stars, $\dot \Omega = \dot J / I$, as a function of their instantaneous velocity. It is seen that the curves for the three mass bins cross-over at intermediate velocity, i.e. higher mass stars are more {rapidly} spun down at high velocities, while lower mass stars are at low velocities. 

\subsection{Gyrochronology and the Skumanich relationship}

Figure \ref{protmass} shows the observed period-mass distributions of the various clusters used in this study together with the model predictions for evolution of slow, median, and fast rotators over the 0.5-1.0~M$_\odot$ mass range. It is seen that the stellar spin smoothly evolves with time at all masses investigated here and models closely follow this evolution. Hence, the possibility of uniquely relating the spin rate to the age of the star, i.e. to explore gyrochronology as originally suggested by \citet{Barnes03} following the pioneering work of \citet{Sku72}. The Skumanich's relationship,  $\Omega_* \propto t^{-1/2}$, was derived for solar-mass stars with an age between 100 Myr and the Sun's. In the models presented here, the asymptotic behaviour of low-mass stars at large ages appear to follow this relationship quite closely (cf. Fig. \ref{model}). It is however important to stress that: i) this relationship is valid only past the time of rotational convergence, i.e. after $\sim$0.5 Gyr for solar-mass stars and after $\sim$3 Gyr for 0.5 M$_\odot$ stars; ii) prior to the epoch of rotational convergence, there is no one-to-one relationship between a star's age and its surface velocity as the initial dispersion of rotation rates has not been erased yet; and iii) the rotational evolution on the early MS can be steeper (fast rotators) or shallower (slow rotators) than the Skumanich relationship would predict depending on the amount of angular momentum being stored in the stellar core and slowly resurfacing into the envelope (cf. Fig. \ref{model}). 

On the late-MS, all stars in the 0.5-1.0 M$_\odot$ mass range have reached rotational convergence. Indeed, the recent results of \citet{Meibom15} on NGC 6819, a 2.5 Gyr open cluster, yield a tight mass-rotation period relationship over the mass range 0.85-1.3~M$_\odot$ at this age. In this cluster, solar-mass stars have rotational periods narrowly distributed within P$_{1M_\odot}$=17-19d; while 0.85 M$_\odot$ stars have periods in the range P$_{0.85M_\odot}$=21-24d. Our solar-mass model interpolated at this age predicts $P_{1M_\odot}$=18.5$\pm$0.8d (with a median period of 17.7d), while the 0.8~M$_\odot$ model yields P$_{0.8M_\odot}$=24.8$\pm$2.0d (median P=22.67d), in excellent agreement with observations. 

The largest set of rotational periods has been recently obtained by \citet{McQuillan2014} for a sample of more than 34000 field stars using the Kepler satellite. We reproduce their results in Fig.~\ref{kepler}  where the spin rates are plotted as a function of mass,  and where we over plotted isochrones predicted from our 0.5, 0.8, and 1.0~M$_\odot$ models. It is seen that the lower envelope of the Kepler rotational distribution is well reproduced by the models at an age of 7 Gyr, as expected for the oldest field stars of the Kepler sample. The upper envelope of the main rotational locus roughly corresponds to 0.5 Gyr-old stars, and the "rain" of stars above it consists of even younger field dwarfs.   

\begin{figure*}[ht!]
     \begin{center}
    \includegraphics[angle=-90,width=1.\textwidth]{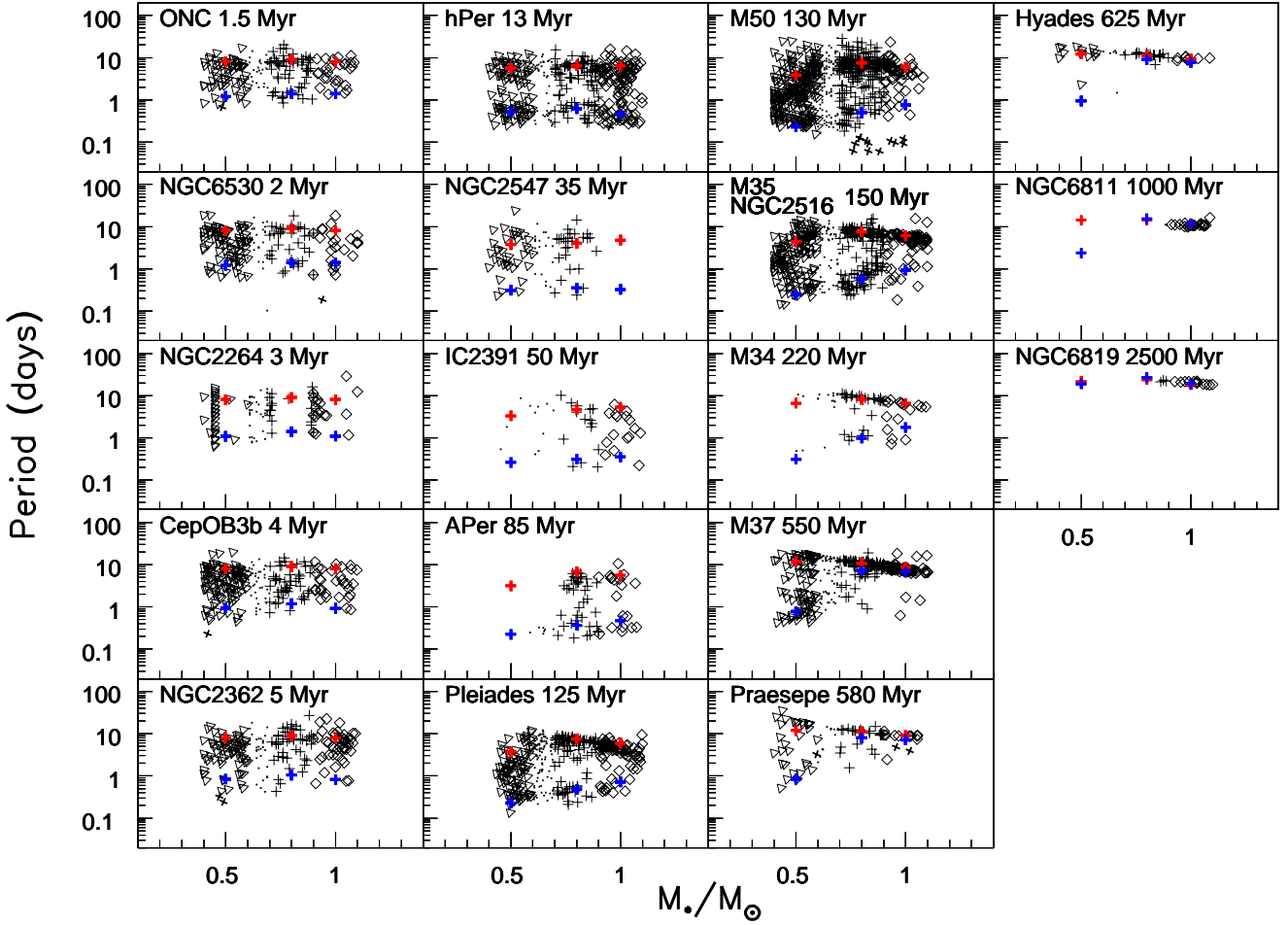}
    \end{center}
    \caption{Same as Fig. \ref{protmass}. The red and blue crosses are the values derived from our evolution model for the slow and fast rotator models, respectively. } %
   \label{protmassmod}
\end{figure*}

\begin{figure*}[ht!]
     \begin{center}
    \includegraphics[angle=-90,width=17cm]{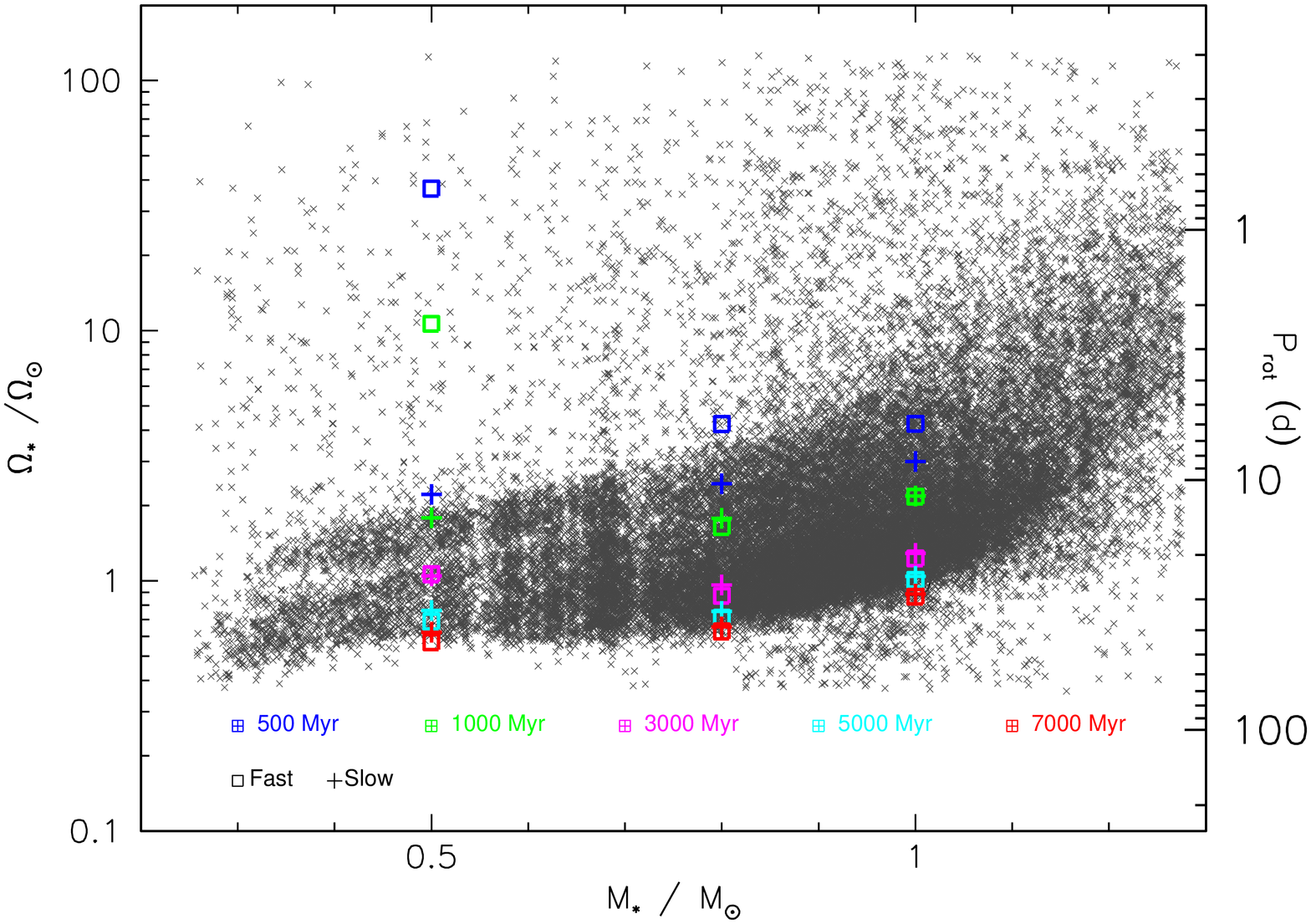}
    \end{center}
    \caption{Angular velocity (left axis) and rotation periods (right axis) as a function of mass for the 34030 field stars whose rotational periods have been measured by \citet{McQuillan2014}.  The stellar mass was derived from $T_{eff}$ using the isochrones of \citet{Baraffe98} for a mean age of 1 Gyr.}%
   \label{kepler}
\end{figure*}

Finally, we should emphasise that there is no dichotomy in the rotational evolution of low-mass stars. We illustrated here models for slow, median, and fast rotators and all three models rely on the same assumptions. Indeed, the evolution of slow and fast rotators alike is driven by the same underlying physical processes. These processes act on different timescales depending on stellar mass and initial velocity, which leads to the observed, peculiar and rapidly evolving, shape of the cluster's rotational distributions (cf. Fig.~\ref{protmassmod}) and, ultimately, to the field distribution (cf. Fig.~\ref{kepler}). Thus, the observed rotational distributions of young open clusters and field stars do not prompt for different processes operating in fast and slow rotators. Instead, the sequence of rotational distributions from 1 Myr to several Gyr can be described as deriving from the smooth temporal evolution of a single, widely dispersed initial distribution of angular momenta at birth. 

{There are others two-zone models in the literature, such as \citet{Keppens95}, \citet{Irwin09a}, and \citet{Den102}. These models are based on the \citet{Kawaler88} braking law and their numerical structures are quite similar to our model (star-disk interaction, core-envelope decoupling, and wind braking). The main difference lies in the wind braking law used. The \citet{Matt12} braking law contains more physics than Kawaler's, especially through the mass-loss rate, which is not included in \citet{Kawaler88}. Moreover, while in these models the saturation of the braking low has to be manually set, in the model presented in this article the saturation is fixed by the observations via the magnetic filling factor.}

{We also highlight that there are other modelling approaches that do not necessarily invoke the two-zone formalism. These models usually assume solid body rotation for the whole stars. \citet{Barnes102} and \citet{Barnes10} propose a model based on gyrochronology analysis \citep{Barnes03} and a Kawaler-modified braking law. This model only works for moderate-to-old MS stars once the convergence is achieved, but fails to reproduce the rotation behaviour of young stars. In \citet{Reiners2012} the braking torque is inversely proportional to the mass-loss rate, which is assumed to be constant during all the evolution. This behaviour somehow contrasts with what we expect, i.e. that the magnetic braking is induced by the quantity of angular momentum extracted by the stellar wind. However, their model can easily be extended to very low-mass star thanks to their mass and radius dependencies. Finally, \citet{Brown2014} proposed a more sophisticated \citet{Barnes102} like model. His models are quite good for intermediate ages ($\approx$ 200 Myr) but fail to reproduce the youngest clusters ($\alpha$ Per) because of a lack of slow rotation (due to the solid body rotation) and the old cluster (M37, NGC6811) especially for low-mass stars.}

\section{Conclusion}
\label{conc}
 
The rotational evolution of low-mass stars in the range from 0.5 to 1.1~M$_{\odot}$ can be described, from birth to the end of the main sequence by parametric models that rely on a limited number of physical processes: star-disk interaction, wind braking, and core-envelope decoupling. The physical processes involved in the angular momentum evolution are included in the models through parametric prescriptions that either rely on observational evidence (e.g. rotational regulation during PMS star-disk interaction), are physically driven simplifications of actual processes  (e.g. core-envelope decoupling), or are based on recent numerical simulations (e.g. wind braking). We explored here how the model parameters vary with mass to get deeper insight into the underlying physical mechanisms. The disk lifetimes and initial rotation periods seem to have little dependence on mass, and the former are consistent with the distribution of disk lifetimes derived from the evolution of IR excess in young stars. However, we do find a correlation between initial period and disk lifetime, which may point to the impact of protostellar disks in establishing the initial distribution of angular momenta. In contrast, the core-envelope coupling timescale and the wind braking efficiency strongly vary with mass, and both increase towards lower mass stars. While the former result may reflect the properties of a still to be identified angular momentum transport process operating in stellar interiors, which is also shown to strongly depend on rotation, the latter may derive from a change in the magnetic topology of dynamo fields as one goes to lower masses. We find that these models do reproduce the run of rotational period distributions as a function of time, from the youngest star-forming regions to the oldest open clusters, and provide a remarkable fit to the oldest field stars from the Kepler sample. The implications are manifold. A large amount of angular momentum must be stored in the radiative core during several 100 Myr on the early main sequence, which will undoubtedly impact on the stellar properties, such as lithium content, even long after the end of the spin-down phase on the main sequence. Also, the build-up of a wide dispersion of rotational velocities at ZAMS from an initially dispersed PMS distribution and its subsequent evolution on the early MS partly reflect this process. The models also naturally account for the longer spin-down timescale of lower mass stars and their lower velocities compared to solar-type stars at the end of the spin-down phase. Finally, we show that the models naturally yield a Skumanich-type rotational evolution on the late main sequence for stars in this mass range, but that until about 1 Gyr, there is no one-to-one relationship between stellar rotation and stellar age, thus undermining the determination of accurate stellar ages for individual young stars from their rotational properties. Indeed, the slow release on the early main sequence of angular momentum hidden in the stellar core delays the epoch at which a Skumanich-type rotational evolution eventually sets in.

\begin{acknowledgements}
This study was supported by the grant ANR 2011 Blanc SIMI5-6 020 01 ``Toupies: Towards understanding the spin evolution of stars" (\url{http://ipag.osug.fr/Anr_Toupies/}). We thank our partners in the ANR project, especially L. Amard for providing us with low-mass star internal parameters. We thank the anonymous referee for helpful comments. We acknowledge financial support from CNRS-INSU's Programme National de Physique Stellaire. 
  
\end{acknowledgements}

%
%
%


\def\jnl@style{\it}
\def\aaref@jnl#1{{\jnl@style#1}}

\def\aaref@jnl#1{{\jnl@style#1}}

\def\aj{\aaref@jnl{AJ}}                   
\def\araa{\aaref@jnl{ARA\&A}}             
\def\apj{\aaref@jnl{ApJ}}                 
\def\apjl{\aaref@jnl{ApJ}}                
\def\apjs{\aaref@jnl{ApJS}}               
\def\ao{\aaref@jnl{Appl.~Opt.}}           
\def\apss{\aaref@jnl{Ap\&SS}}             
\def\aap{\aaref@jnl{A\&A}}                
\def\aapr{\aaref@jnl{A\&A~Rev.}}          
\def\aaps{\aaref@jnl{A\&AS}}              
\def\azh{\aaref@jnl{AZh}}                 
\def\baas{\aaref@jnl{BAAS}}               
\def\jrasc{\aaref@jnl{JRASC}}             
\def\memras{\aaref@jnl{MmRAS}}            
\def\mnras{\aaref@jnl{MNRAS}}             
\def\pra{\aaref@jnl{Phys.~Rev.~A}}        
\def\prb{\aaref@jnl{Phys.~Rev.~B}}        
\def\prc{\aaref@jnl{Phys.~Rev.~C}}        
\def\prd{\aaref@jnl{Phys.~Rev.~D}}        
\def\pre{\aaref@jnl{Phys.~Rev.~E}}        
\def\prl{\aaref@jnl{Phys.~Rev.~Lett.}}    
\def\pasp{\aaref@jnl{PASP}}               
\def\pasj{\aaref@jnl{PASJ}}               
\def\qjras{\aaref@jnl{QJRAS}}             
\def\skytel{\aaref@jnl{S\&T}}             
\def\solphys{\aaref@jnl{Sol.~Phys.}}      
\def\sovast{\aaref@jnl{Soviet~Ast.}}      
\def\ssr{\aaref@jnl{Space~Sci.~Rev.}}     
\def\zap{\aaref@jnl{ZAp}}                 
\def\nat{\aaref@jnl{Nature}}              
\def\iaucirc{\aaref@jnl{IAU~Circ.}}       
\def\aplett{\aaref@jnl{Astrophys.~Lett.}} 
\def\apspr{\aaref@jnl{Astrophys.~Space~Phys.~Res.}}
\def\bain{\aaref@jnl{Bull.~Astron.~Inst.~Netherlands}} 
\def\fcp{\aaref@jnl{Fund.~Cosmic~Phys.}}  
\def\gca{\aaref@jnl{Geochim.~Cosmochim.~Acta}}   
\def\grl{\aaref@jnl{Geophys.~Res.~Lett.}} 
\def\jcp{\aaref@jnl{J.~Chem.~Phys.}}      
\def\jgr{\aaref@jnl{J.~Geophys.~Res.}}    
\def\jqsrt{\aaref@jnl{J.~Quant.~Spec.~Radiat.~Transf.}}
\def\memsai{\aaref@jnl{Mem.~Soc.~Astron.~Italiana}}
\def\nphysa{\aaref@jnl{Nucl.~Phys.~A}}   
\def\physrep{\aaref@jnl{Phys.~Rep.}}   
\def\physscr{\aaref@jnl{Phys.~Scr}}   
\def\planss{\aaref@jnl{Planet.~Space~Sci.}}   
\def\procspie{\aaref@jnl{Proc.~SPIE}}   

\let\astap=\aap
\let\apjlett=\apjl
\let\apjsupp=\apjs
\let\applopt=\ao

\bibliographystyle{aa}
\bibliography{Bib}

\newcommand{\noop}[1]{}
\begin{thebibliography}{91}
\expandafter\ifx\csname natexlab\endcsname\relax\def\natexlab#1{#1}\fi

\bibitem[{{Affer} {et~al.}(2012){Affer}, {Micela}, {Favata}, \&
  {Flaccomio}}]{Affer2012}
{Affer}, L., {Micela}, G., {Favata}, F., \& {Flaccomio}, E. 2012, \mnras, 424,
  11

\bibitem[{{Affer} {et~al.}(2013){Affer}, {Micela}, {Favata}, {Flaccomio}, \&
  {Bouvier}}]{Affer2013}
{Affer}, L., {Micela}, G., {Favata}, F., {Flaccomio}, E., \& {Bouvier}, J.
  2013, ArXiv e-prints

\bibitem[{{Ag{\"u}eros} {et~al.}(2011){Ag{\"u}eros}, {Covey}, {Lemonias},
  {Law}, {Kraus}, {Batalha}, {Bloom}, {Cenko}, {Kasliwal}, {Kulkarni},
  {Nugent}, {Ofek}, {Poznanski}, \& {Quimby}}]{Agueros11}
{Ag{\"u}eros}, M.~A., {Covey}, K.~R., {Lemonias}, J.~J., {et~al.} 2011, \apj,
  740, 110

\bibitem[{{Allain}(1998)}]{Allain98}
{Allain}, S. 1998, \aap, 333, 629

\bibitem[{{Baraffe} {et~al.}(1998){Baraffe}, {Chabrier}, {Allard}, \&
  {Hauschildt}}]{Baraffe98}
{Baraffe}, I., {Chabrier}, G., {Allard}, F., \& {Hauschildt}, P.~H. 1998, \aap,
  337, 403

\bibitem[{{Barnes}(2003)}]{Barnes03}
{Barnes}, S.~A. 2003, \apj, 586, 464

\bibitem[{{Barnes}(2010)}]{Barnes102}
{Barnes}, S.~A. 2010, \apj, 722, 222

\bibitem[{{Barnes} \& {Kim}(2010)}]{Barnes10}
{Barnes}, S.~A. \& {Kim}, Y.-C. 2010, \apj, 721, 675

\bibitem[{{Bell} {et~al.}(2013){Bell}, {Naylor}, {Mayne}, {Jeffries}, \&
  {Littlefair}}]{Bell13}
{Bell}, C.~P.~M., {Naylor}, T., {Mayne}, N.~J., {Jeffries}, R.~D., \&
  {Littlefair}, S.~P. 2013, \mnras, 434, 806

\bibitem[{{Bouvier}(2008)}]{Bouvier08}
{Bouvier}, J. 2008, \aap, 489, L53

\bibitem[{{Bouvier} {et~al.}(2013){Bouvier}, {Matt}, {Mohanty}, {Scholz},
  {Stassun}, \& {Zanni}}]{PPVI}
{Bouvier}, J., {Matt}, S.~P., {Mohanty}, S., {et~al.} 2013, ArXiv e-prints

\bibitem[{{Brown}(2014)}]{Brown2014}
{Brown}, T.~M. 2014, \apj, 789, 101

\bibitem[{{Brun} {et~al.}(2011){Brun}, {Miesch}, \& {Toomre}}]{Brun11}
{Brun}, A.~S., {Miesch}, M.~S., \& {Toomre}, J. 2011, \apj, 742, 79

\bibitem[{{Charbonnel} {et~al.}(2013){Charbonnel}, {Decressin}, {Amard},
  {Palacios}, \& {Talon}}]{Charbonnel13}
{Charbonnel}, C., {Decressin}, T., {Amard}, L., {Palacios}, A., \& {Talon}, S.
  2013, \aap, 554, A40

\bibitem[{{Charbonnel} \& {Lagarde}(2010)}]{Charbonnel10}
{Charbonnel}, C. \& {Lagarde}, N. 2010, \aap, 522, A10

\bibitem[{{Charbonnel} \& {Talon}(2005)}]{Charbonnel05}
{Charbonnel}, C. \& {Talon}, S. 2005, Science, 309, 2189

\bibitem[{{Cieza} \& {Baliber}(2007)}]{Cieza07}
{Cieza}, L. \& {Baliber}, N. 2007, \apj, 671, 605

\bibitem[{{Cranmer} \& {Saar}(2011)}]{Cranmer11}
{Cranmer}, S.~R. \& {Saar}, S.~H. 2011, \apj, 741, 54

\bibitem[{{Deheuvels} {et~al.}(2012){Deheuvels}, {Garc{\'{\i}}a}, {Chaplin},
  {Basu}, {Antia}, {Appourchaux}, {Benomar}, {Davies}, {Elsworth}, {Gizon},
  {Goupil}, {Reese}, {Regulo}, {Schou}, {Stahn}, {Casagrande},
  {Christensen-Dalsgaard}, {Fischer}, {Hekker}, {Kjeldsen}, {Mathur}, {Mosser},
  {Pinsonneault}, {Valenti}, {Christiansen}, {Kinemuchi}, \&
  {Mullally}}]{Deheuvels12}
{Deheuvels}, S., {Garc{\'{\i}}a}, R.~A., {Chaplin}, W.~J., {et~al.} 2012, \apj,
  756, 19

\bibitem[{{Delorme} {et~al.}(2011){Delorme}, {Collier Cameron}, {Hebb},
  {Rostron}, {Lister}, {Norton}, {Pollacco}, \& {West}}]{Delorme11}
{Delorme}, P., {Collier Cameron}, A., {Hebb}, L., {et~al.} 2011, \mnras, 413,
  2218

\bibitem[{{Denissenkov}(2010)}]{Den102}
{Denissenkov}, P.~A. 2010, \apj, 719, 28

\bibitem[{{Denissenkov} \& {Pinsonneault}(2007)}]{Den07}
{Denissenkov}, P.~A. \& {Pinsonneault}, M. 2007, \apj, 655, 1157

\bibitem[{{Denissenkov} {et~al.}(2010){Denissenkov}, {Pinsonneault},
  {Terndrup}, \& {Newsham}}]{Den10}
{Denissenkov}, P.~A., {Pinsonneault}, M., {Terndrup}, D.~M., \& {Newsham}, G.
  2010, \apj, 716, 1269

\bibitem[{{Donati} \& {Landstreet}(2009)}]{Donati09}
{Donati}, J.-F. \& {Landstreet}, J.~D. 2009, \araa, 47, 333

\bibitem[{{Eggenberger} {et~al.}(2012{\natexlab{a}}){Eggenberger},
  {Haemmerl{\'e}}, {Meynet}, \& {Maeder}}]{Eggenberger12}
{Eggenberger}, P., {Haemmerl{\'e}}, L., {Meynet}, G., \& {Maeder}, A.
  2012{\natexlab{a}}, \aap, 539, A70

\bibitem[{{Eggenberger} {et~al.}(2012{\natexlab{b}}){Eggenberger},
  {Montalb{\'a}n}, \& {Miglio}}]{Eggenberger12b}
{Eggenberger}, P., {Montalb{\'a}n}, J., \& {Miglio}, A. 2012{\natexlab{b}},
  \aap, 544, L4

\bibitem[{{Ferreira} {et~al.}(2000){Ferreira}, {Pelletier}, \& {Appl}}]{FPA00}
{Ferreira}, J., {Pelletier}, G., \& {Appl}, S. 2000, \mnras, 312, 387

\bibitem[{{Gallet} \& {Bouvier}(2013)}]{Gallet13}
{Gallet}, F. \& {Bouvier}, J. 2013, \aap, 556, A36

\bibitem[{{Gregory} {et~al.}(2012){Gregory}, {Donati}, {Morin}, {Hussain},
  {Mayne}, {Hillenbrand}, \& {Jardine}}]{Gregory2012}
{Gregory}, S.~G., {Donati}, J.-F., {Morin}, J., {et~al.} 2012, \apj, 755, 97

\bibitem[{{Hartman} {et~al.}(2010){Hartman}, {Bakos}, {Kov{\'a}cs}, \&
  {Noyes}}]{Hartman10}
{Hartman}, J.~D., {Bakos}, G.~{\'A}., {Kov{\'a}cs}, G., \& {Noyes}, R.~W. 2010,
  \mnras, 408, 475

\bibitem[{{Hartman} {et~al.}(2009){Hartman}, {Gaudi}, {Pinsonneault}, {Stanek},
  {Holman}, {McLeod}, {Meibom}, {Barranco}, \& {Kalirai}}]{Hartman09}
{Hartman}, J.~D., {Gaudi}, B.~S., {Pinsonneault}, M.~H., {et~al.} 2009, \apj,
  691, 342

\bibitem[{{Heger} {et~al.}(2000){Heger}, {Langer}, \& {Woosley}}]{Heger00}
{Heger}, A., {Langer}, N., \& {Woosley}, S.~E. 2000, \apj, 528, 368

\bibitem[{{Henderson} \& {Stassun}(2012)}]{Henderson11}
{Henderson}, C.~B. \& {Stassun}, K.~G. 2012, \apj, 747, 51

\bibitem[{{Hern{\'a}ndez} {et~al.}(2008){Hern{\'a}ndez}, {Hartmann}, {Calvet},
  {Jeffries}, {Gutermuth}, {Muzerolle}, \& {Stauffer}}]{Hernandez08}
{Hern{\'a}ndez}, J., {Hartmann}, L., {Calvet}, N., {et~al.} 2008, \apj, 686,
  1195

\bibitem[{{Hillenbrand}(1997)}]{Hillenbrand97}
{Hillenbrand}, L.~A. 1997, \aj, 113, 1733

\bibitem[{{Irwin} {et~al.}(2009){Irwin}, {Aigrain}, {Bouvier}, {Hebb},
  {Hodgkin}, {Irwin}, \& {Moraux}}]{Irwin09b}
{Irwin}, J., {Aigrain}, S., {Bouvier}, J., {et~al.} 2009, \mnras, 392, 1456

\bibitem[{{Irwin} {et~al.}(2011){Irwin}, {Berta}, {Burke}, {Charbonneau},
  {Nutzman}, {West}, \& {Falco}}]{Irwin11}
{Irwin}, J., {Berta}, Z.~K., {Burke}, C.~J., {et~al.} 2011, \apj, 727, 56

\bibitem[{{Irwin} \& {Bouvier}(2009)}]{Irwin09a}
{Irwin}, J. \& {Bouvier}, J. 2009, in IAU Symposium, Vol. 258, IAU Symposium,
  ed. {E.~E.~Mamajek, D.~R.~Soderblom, \& R.~F.~G.~Wyse}, 363--374

\bibitem[{{Irwin} {et~al.}(2008{\natexlab{a}}){Irwin}, {Hodgkin}, {Aigrain},
  {Bouvier}, {Hebb}, {Irwin}, \& {Moraux}}]{Irwin08a}
{Irwin}, J., {Hodgkin}, S., {Aigrain}, S., {et~al.} 2008{\natexlab{a}}, \mnras,
  384, 675

\bibitem[{{Irwin} {et~al.}(2008{\natexlab{b}}){Irwin}, {Hodgkin}, {Aigrain},
  {Bouvier}, {Hebb}, \& {Moraux}}]{Irwin08b}
{Irwin}, J., {Hodgkin}, S., {Aigrain}, S., {et~al.} 2008{\natexlab{b}}, \mnras,
  383, 1588

\bibitem[{{Irwin} {et~al.}(2007){Irwin}, {Hodgkin}, {Aigrain}, {Hebb},
  {Bouvier}, {Clarke}, {Moraux}, \& {Bramich}}]{Irwin07}
{Irwin}, J., {Hodgkin}, S., {Aigrain}, S., {et~al.} 2007, \mnras, 377, 741

\bibitem[{{Kawaler}(1988)}]{Kawaler88}
{Kawaler}, S.~D. 1988, \apj, 333, 236

\bibitem[{{Keppens} {et~al.}(1995){Keppens}, {MacGregor}, \&
  {Charbonneau}}]{Keppens95}
{Keppens}, R., {MacGregor}, K.~B., \& {Charbonneau}, P. 1995, \aap, 294, 469

\bibitem[{{Krishnamurthi} {et~al.}(1997){Krishnamurthi}, {Pinsonneault},
  {Barnes}, \& {Sofia}}]{Kri97}
{Krishnamurthi}, A., {Pinsonneault}, M.~H., {Barnes}, S., \& {Sofia}, S. 1997,
  \apj, 480, 303

\bibitem[{{Lagarde} {et~al.}(2011){Lagarde}, {Charbonnel}, {Decressin}, \&
  {Hagelberg}}]{Lagarde11}
{Lagarde}, N., {Charbonnel}, C., {Decressin}, T., \& {Hagelberg}, J. 2011,
  \aap, 536, A28

\bibitem[{{Lagarde} {et~al.}(2012){Lagarde}, {Decressin}, {Charbonnel},
  {Eggenberger}, {Ekstr{\"o}m}, \& {Palacios}}]{Lagarde12}
{Lagarde}, N., {Decressin}, T., {Charbonnel}, C., {et~al.} 2012, \aap, 543,
  A108

\bibitem[{{Littlefair} {et~al.}(2010){Littlefair}, {Naylor}, {Mayne},
  {Saunders}, \& {Jeffries}}]{Littlefair10}
{Littlefair}, S.~P., {Naylor}, T., {Mayne}, N.~J., {Saunders}, E.~S., \&
  {Jeffries}, R.~D. 2010, \mnras, 403, 545

\bibitem[{{MacGregor} \& {Brenner}(1991)}]{McGB91}
{MacGregor}, K.~B. \& {Brenner}, M. 1991, \apj, 376, 204

\bibitem[{{Matt} \& {Pudritz}(2005{\natexlab{a}})}]{MP05b}
{Matt}, S. \& {Pudritz}, R.~E. 2005{\natexlab{a}}, \apjl, 632, L135

\bibitem[{{Matt} \& {Pudritz}(2005{\natexlab{b}})}]{MP05a}
{Matt}, S. \& {Pudritz}, R.~E. 2005{\natexlab{b}}, \mnras, 356, 167

\bibitem[{{Matt} \& {Pudritz}(2008{\natexlab{a}})}]{MP08a}
{Matt}, S. \& {Pudritz}, R.~E. 2008{\natexlab{a}}, \apj, 678, 1109

\bibitem[{{Matt} \& {Pudritz}(2008{\natexlab{b}})}]{MP08b}
{Matt}, S. \& {Pudritz}, R.~E. 2008{\natexlab{b}}, \apj, 681, 391

\bibitem[{{Matt} {et~al.}(2014){Matt}, {Brun}, {Baraffe}, {Bouvier}, \&
  {Chabrier}}]{Matt2014}
{Matt}, S.~P., {Brun}, A.~S., {Baraffe}, I., {Bouvier}, J., \& {Chabrier}, G.
  2014, ArXiv e-prints

\bibitem[{{Matt} {et~al.}(2012{\natexlab{a}}){Matt}, {MacGregor},
  {Pinsonneault}, \& {Greene}}]{Matt12}
{Matt}, S.~P., {MacGregor}, K.~B., {Pinsonneault}, M.~H., \& {Greene}, T.~P.
  2012{\natexlab{a}}, \apjl, 754, L26

\bibitem[{{Matt} {et~al.}(2010){Matt}, {Pinz{\'o}n}, {de la Reza}, \&
  {Greene}}]{Matt10}
{Matt}, S.~P., {Pinz{\'o}n}, G., {de la Reza}, R., \& {Greene}, T.~P. 2010,
  \apj, 714, 989

\bibitem[{{Matt} {et~al.}(2012{\natexlab{b}}){Matt}, {Pinz{\'o}n}, {Greene}, \&
  {Pudritz}}]{Matt11a}
{Matt}, S.~P., {Pinz{\'o}n}, G., {Greene}, T.~P., \& {Pudritz}, R.~E.
  2012{\natexlab{b}}, \apj, 745, 101

\bibitem[{{McQuillan} {et~al.}(2013){McQuillan}, {Aigrain}, \&
  {Mazeh}}]{McQuillan2013}
{McQuillan}, A., {Aigrain}, S., \& {Mazeh}, T. 2013, \mnras, 432, 1203

\bibitem[{{McQuillan} {et~al.}(2014){McQuillan}, {Mazeh}, \&
  {Aigrain}}]{McQuillan2014}
{McQuillan}, A., {Mazeh}, T., \& {Aigrain}, S. 2014, \apjs, 211, 24

\bibitem[{{Meibom} {et~al.}(2011{\natexlab{a}}){Meibom}, {Barnes}, {Latham},
  {Batalha}, {Borucki}, {Koch}, {Basri}, {Walkowicz}, {Janes}, {Jenkins}, {Van
  Cleve}, {Haas}, {Bryson}, {Dupree}, {Furesz}, {Szentgyorgyi}, {Buchhave},
  {Clarke}, {Twicken}, \& {Quintana}}]{Meibom2011}
{Meibom}, S., {Barnes}, S.~A., {Latham}, D.~W., {et~al.} 2011{\natexlab{a}},
  \apjl, 733, L9

\bibitem[{{Meibom} {et~al.}(2015){Meibom}, {Barnes}, {Platais}, {Gilliland},
  {Latham}, \& {Mathieu}}]{Meibom15}
{Meibom}, S., {Barnes}, S.~A., {Platais}, I., {et~al.} 2015, \nat, 517, 589

\bibitem[{{Meibom} {et~al.}(2009){Meibom}, {Mathieu}, \& {Stassun}}]{Meibom09}
{Meibom}, S., {Mathieu}, R.~D., \& {Stassun}, K.~G. 2009, \apj, 695, 679

\bibitem[{{Meibom} {et~al.}(2011{\natexlab{b}}){Meibom}, {Mathieu}, {Stassun},
  {Liebesny}, \& {Saar}}]{Meibom2011b}
{Meibom}, S., {Mathieu}, R.~D., {Stassun}, K.~G., {Liebesny}, P., \& {Saar},
  S.~H. 2011{\natexlab{b}}, \apj, 733, 115

\bibitem[{{Moraux} {et~al.}(2013){Moraux}, {Artemenko}, {Bouvier}, {Irwin},
  {Ibrahimov}, {Magakian}, {Grankin}, {Nikogossian}, {Cardoso}, {Hodgkin},
  {Aigrain}, \& {Movsessian}}]{Moraux13}
{Moraux}, E., {Artemenko}, S., {Bouvier}, J., {et~al.} 2013, ArXiv e-prints

\bibitem[{{Morin} {et~al.}(2010){Morin}, {Donati}, {Petit}, {Delfosse},
  {Forveille}, \& {Jardine}}]{Morin10}
{Morin}, J., {Donati}, J.-F., {Petit}, P., {et~al.} 2010, \mnras, 407, 2269

\bibitem[{{Mosser} {et~al.}(2012){Mosser}, {Goupil}, {Belkacem}, {Marques},
  {Beck}, {Bloemen}, {De Ridder}, {Barban}, {Deheuvels}, {Elsworth}, {Hekker},
  {Kallinger}, {Ouazzani}, {Pinsonneault}, {Samadi}, {Stello}, {Garc{\'{\i}}a},
  {Klaus}, {Li}, {Mathur}, \& {Morris}}]{Mosser12}
{Mosser}, B., {Goupil}, M.~J., {Belkacem}, K., {et~al.} 2012, \aap, 548, A10

\bibitem[{{Oglethorpe} \& {Garaud}(2013)}]{Oglethorpe13}
{Oglethorpe}, R.~L.~F. \& {Garaud}, P. 2013, \apj, 778, 166

\bibitem[{{Palacios} {et~al.}(2006){Palacios}, {Charbonnel}, {Talon}, \&
  {Siess}}]{Palacios06}
{Palacios}, A., {Charbonnel}, C., {Talon}, S., \& {Siess}, L. 2006, \aap, 453,
  261

\bibitem[{{Palacios} {et~al.}(2003){Palacios}, {Talon}, {Charbonnel}, \&
  {Forestini}}]{Palacios03}
{Palacios}, A., {Talon}, S., {Charbonnel}, C., \& {Forestini}, M. 2003, \aap,
  399, 603

\bibitem[{{Petit} {et~al.}(2008){Petit}, {Dintrans}, {Solanki}, {Donati},
  {Auri{\`e}re}, {Ligni{\`e}res}, {Morin}, {Paletou}, {Ramirez Velez},
  {Catala}, \& {Fares}}]{Petit08}
{Petit}, P., {Dintrans}, B., {Solanki}, S.~K., {et~al.} 2008, \mnras, 388, 80

\bibitem[{{Reiners} \& {Mohanty}(2012)}]{Reiners2012}
{Reiners}, A. \& {Mohanty}, S. 2012, \apj, 746, 43

\bibitem[{{R{\'e}ville} {et~al.}(2015){R{\'e}ville}, {Brun}, {Matt},
  {Strugarek}, \& {Pinto}}]{Reville14}
{R{\'e}ville}, V., {Brun}, A.~S., {Matt}, S.~P., {Strugarek}, A., \& {Pinto},
  R.~F. 2015, \apj, 798, 116

\bibitem[{{Rodr{\'{\i}}guez-Ledesma} {et~al.}(2009){Rodr{\'{\i}}guez-Ledesma},
  {Mundt}, \& {Eisl{\"o}ffel}}]{RL2009}
{Rodr{\'{\i}}guez-Ledesma}, M.~V., {Mundt}, R., \& {Eisl{\"o}ffel}, J. 2009,
  \aap, 502, 883

\bibitem[{{Siess} {et~al.}(2000){Siess}, {Dufour}, \& {Forestini}}]{Siess00}
{Siess}, L., {Dufour}, E., \& {Forestini}, M. 2000, \aap, 358, 593

\bibitem[{{Skumanich}(1972)}]{Sku72}
{Skumanich}, A. 1972, \apj, 171, 565

\bibitem[{{Spada} {et~al.}(2010){Spada}, {Lanzafame}, \& {Lanza}}]{Spada10}
{Spada}, F., {Lanzafame}, A.~C., \& {Lanza}, A.~F. 2010, \mnras, 404, 641

\bibitem[{{Spada} {et~al.}(2011){Spada}, {Lanzafame}, {Lanza}, {Messina}, \&
  {Collier Cameron}}]{Spada11}
{Spada}, F., {Lanzafame}, A.~C., {Lanza}, A.~F., {Messina}, S., \& {Collier
  Cameron}, A. 2011, \mnras, 416, 447

\bibitem[{{Talon} \& {Charbonnel}(2003)}]{Talon03}
{Talon}, S. \& {Charbonnel}, C. 2003, \aap, 405, 1025

\bibitem[{{Talon} \& {Charbonnel}(2005)}]{Talon05}
{Talon}, S. \& {Charbonnel}, C. 2005, \aap, 440, 981

\bibitem[{{Turck-Chieze} {et~al.}(2011){Turck-Chieze}, {Couvidat},
  {Eff-Darwich}, {Duez}, {Garcia}, {Mathis}, {Mathur}, {Piau}, \&
  {Salabert}}]{TC2011}
{Turck-Chieze}, S., {Couvidat}, S., {Eff-Darwich}, A., {et~al.} 2011, ArXiv
  e-prints

\bibitem[{{Vidotto} {et~al.}(2014{\natexlab{a}}){Vidotto}, {Gregory},
  {Jardine}, {Donati}, {Petit}, {Morin}, {Folsom}, {Bouvier}, {Cameron},
  {Hussain}, {Marsden}, {Waite}, {Fares}, {Jeffers}, \& {do
  Nascimento}}]{Vidotto2014}
{Vidotto}, A.~A., {Gregory}, S.~G., {Jardine}, M., {et~al.} 2014{\natexlab{a}},
  ArXiv e-prints

\bibitem[{{Vidotto} {et~al.}(2013){Vidotto}, {Jardine}, {Morin}, {Donati},
  {Opher}, \& {Gombosi}}]{Vidotto2013}
{Vidotto}, A.~A., {Jardine}, M., {Morin}, J., {et~al.} 2013, \mnras

\bibitem[{{Vidotto} {et~al.}(2014{\natexlab{b}}){Vidotto}, {Jardine}, {Morin},
  {Donati}, {Opher}, \& {Gombosi}}]{Vidotto14b}
{Vidotto}, A.~A., {Jardine}, M., {Morin}, J., {et~al.} 2014{\natexlab{b}},
  \mnras, 438, 1162

\bibitem[{{Vidotto} {et~al.}(2011){Vidotto}, {Jardine}, {Opher}, {Donati}, \&
  {Gombosi}}]{Vidotto2011}
{Vidotto}, A.~A., {Jardine}, M., {Opher}, M., {Donati}, J.~F., \& {Gombosi},
  T.~I. 2011, \mnras, 412, 351

\bibitem[{{Weber} \& {Davis}(1967)}]{WD67}
{Weber}, E.~J. \& {Davis}, Jr., L. 1967, \apj, 148, 217

\bibitem[{{Williams} \& {Cieza}(2011)}]{Williams11}
{Williams}, J.~P. \& {Cieza}, L.~A. 2011, \araa, 49, 67

\bibitem[{{Wright} {et~al.}(2011){Wright}, {Drake}, {Mamajek}, \&
  {Henry}}]{Wright11}
{Wright}, N.~J., {Drake}, J.~J., {Mamajek}, E.~E., \& {Henry}, G.~W. 2011,
  \apj, 743, 48

\bibitem[{{Wyatt}(2008)}]{Wyatt08}
{Wyatt}, M.~C. 2008, \araa, 46, 339

\bibitem[{{Zahn} {et~al.}(1997){Zahn}, {Talon}, \& {Matias}}]{Zahn97}
{Zahn}, J.-P., {Talon}, S., \& {Matias}, J. 1997, \aap, 322, 320

\bibitem[{{Zanni} \& {Ferreira}(2009)}]{Zanni09}
{Zanni}, C. \& {Ferreira}, J. 2009, \aap, 508, 1117

\bibitem[{{Zanni} \& {Ferreira}(2011)}]{Zanni11}
{Zanni}, C. \& {Ferreira}, J. 2011, \apjl, 727, L22+

\bibitem[{{Zanni} \& {Ferreira}(2013)}]{Zanni2012}
{Zanni}, C. \& {Ferreira}, J. 2013, \aap, 550, A99

\end{thebibliography}

\end{document}